\documentclass[review]{elsarticle}

\usepackage{lineno,hyperref}
\usepackage{graphicx}
\usepackage{color}

\modulolinenumbers[5]

\journal{Journal of \LaTeX\ Templates}

\begin{document}

\begin{frontmatter}

\title{Development  and application of CATCH: A cylindrical active tracker and calorimeter system for hyperon--proton scattering experiments}


\author{Y. Akazawa$^{1}$}
\author{N. Chiga$^2$}
\author{N. Fujioka$^2$}
\author{S.H. Hayakawa$^3$}
\author{R. Honda$^1$}
\author{M. Ikeda$^2$}
\author{K. Matsuda$^2$}
\author{K. Miwa$^2$\corref{mycorrespondingauthor}}
\cortext[mycorrespondingauthor]{Corresponding author}
\ead{miwa9@lambda.phys.tohoku.ac.jp}
\author{Y. Nakada$^4$}
\author{T. Nanamura$^{3,5}$}
\author{S. Ozawa$^2$}
\author{T. Shiozaki$^2$}
\author{H. Tamura$^{2,3}$}
\author{H. Umetsu$^2$}

\address{$^1$ High Energy Accelerator Research Organization (KEK), Japan}
\address{$^2$ Department of Physics, Tohoku University, Japan}
\address{$^3$ Japan Atomic Energy Agency (JAEA), Japan}
\address{$^4$ Department of Physics, Osaka University, Japan}
\address{$^5$ Department of Physics, Kyoto University, Japan}

\begin{abstract}
In this study, we  developed a new proton detector system called CATCH, which was designed for a scattering experiment involving a $\Sigma$ hyperon and a proton (J-PARC E40). 
CATCH is a cylindrical detector system covering an inner target that can be used to measure the trajectory and energy of a proton emitted from the target for the kinematic identification of a $\Sigma p$ scattering event.
It  comprises a cylindrical fiber tracker (CFT), a bismuth germanate (BGO) calorimeter, and a plastic scintillator hodoscope (PiID), which are coaxially arranged from the inner to the outer sides. 
The CFT is a tracking detector consisting of 5,000 scintillation fibers, and it has 
 two types of cylindrical layers in which the fibers are arranged in straight and spiral configurations.  
 24 BGO crystals are placed around the CFT to measure the kinetic energy of the recoil proton. The PiID is used to determine whether the recoil proton is stopped in the BGO calorimeter .
We performed proton-- proton ($pp$) and proton--carbon ($p$C) scattering experiments using an 80 MeV proton beam to evaluate the performance of CATCH. 
The total energy resolution for the recoil proton was 2.8 MeV in $\sigma$ for the entire angular region after the energy calibrations of the BGO calorimeter and the CFT. The angular resolution of the CFT was 1.27 degrees in $\sigma$ for the proton, and the time resolution was more than 1.8 ns in $\sigma$.
We also developed an analysis method for deriving the cross section of the $pp$ scattering using CATCH.
The obtained relative differential cross section for the $pp$ elastic scattering was consistent with that obtained by reliable partial wave analysis, and the systematic error was maintained at below 10\%.  These performance results satisfy our requirements for a reliable detection system for the $\Sigma p$ scattering experiment conducted at J-PARC.

\end{abstract}


\end{frontmatter}


\section{Introduction} 

Scattering experiments involving a hyperon and a proton are the most effective methods for investigating two-body hyperon--nucleon ($YN$) interactions, as is the case in various intensive studies on $pp$ and $np$ \color{black} scattering \color{black}, which are aimed at understanding nucleon--nucleon \color{black} ($NN$) \color{black} interactions.  Scattering observables, such as differential cross sections and spin observables, are essential experimental inputs for constructing theoretical frameworks of $YN$ interactions based on the one-boson-exchange model assuming a broken flavor SU(3) symmetry \cite{1, Rijken:1999, Nagels:2019, Haidenbauer:2005}.
The introduction of strange quarks is essential for studying the role of quarks in the short-range region, where two baryons overlap with each other.
This is because the spin-flavor SU(6) quark model  predict significantly characteristic potentials in the short-range region, such as significantly repulsive cores or attractive interactions depending on the spin and flavor configuration of the quarks in the system \cite{Oka:1986, Fujiwara:2007, Inoue:2012}.
However, regarding hyperon-proton scattering experiments, no experimental progress has been made from the experiments conducted throughout the 1970s in which hydrogen bubble chambers were employed \cite{Sechi-zorn:1968, Alexander:1968, Kadyk:1971, Hauptman:1977, Engelmann:1966, Eisele:1971, Stephen:1970}. This is owing to the experimental difficulties stemming from the low intensity of the hyperon beam and its short lifetime. 

\color{black}
A new hyperon--proton scattering experiment, dubbed J-PARC E40, was proposed to measure differential cross sections of the $\Sigma^{+}p$, $\Sigma^{-}p$ \cite{Miwa:2021} elastic scatterings and the $\Sigma^{-}p \to \Lambda n$ scattering \cite{Miwa:2021_2} by detecting more than 10,000 scattering events in the $\Sigma$ momentum region ranging from 0.4 to 0.8 GeV/$c$.
\color{black}
This experiment consists of two detector components.
The first component is the magnetic spectrometers  for identifying $\Sigma$ production.
Because the $\Sigma$ hyperon's lifetime is too short to be delivered through a normal beam line, the $\Sigma$ hyperon is produced inside a liquid hydrogen (LH$_{2}$) target through $\pi^{\pm}p \to K^{+}\Sigma^{\pm}$ reactions, and the momentum vector of the produced $\Sigma$ is tagged from the missing momentum in the $\pi^{\pm}p \to K^{+}X$ reaction measured using the magnetic spectrometers.
We regard such a $\Sigma$ hyperon running in the LH$_{2}$ target as a $\Sigma$ hyperon beam.
The $\Sigma$ hyperon interacts with a proton in the LH$_{2}$ target with a probability that depends on the interaction cross section between a $\Sigma$ hyperon and a proton.
The second component comprises a detector system that is placed around the LH$_{2}$ target to identify the $\Sigma p$ scattering by detecting the recoil proton and the decay products resulting from the $\Sigma$ decay.
To identify the $\Sigma p$ scattering events, the kinetic energy and the recoil angle of the recoil proton should be measured to determine whether the detected proton is consistent with the $\Sigma p$ scattering kinematics.
For this purpose, we developed a dedicated detector system called CATCH, which comprises a cylindrical fiber tracker (CFT) for tracking, a BGO calorimeter for kinetic energy measurement, and a plastic scintillator hodoscope (PiID) for providing additional information regarding particle differentiation  between a proton and $\pi$ by verifying whether the particle is stopped in the BGO calorimeter.

In this paper, we describe the development and performance evaluation of the newly constructed system called CATCH, including basic points, such as the design concept, experimental requirements, and the fabrication of the detector system.


\section{Overview of CATCH for a $\Sigma p$ scattering experiment }

\begin{figure}[]
\centering
\includegraphics[width=80mm]{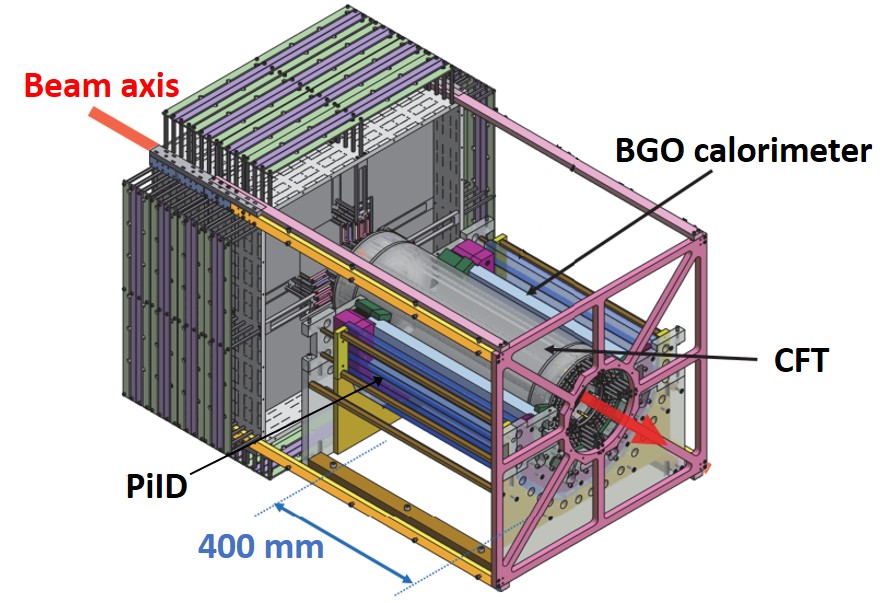}
\caption[CATCH]{
Proposed detector system called CATCH.
CATCH comprises a cylindrical fiber tracker (CFT) and a BGO calorimeter for measuring the trajectory and energy of the protons emitted from the target installed inside CATCH. 
The PiID is used to determine whether the recoil proton is stopped in the BGO calorimeter. 
The upper half of the BGO calorimeter and the PiID is omitted in this figure.
}
\label{catch}
\end{figure}

CATCH is a cylindrical detector system for detecting recoil protons and decay products from $\Sigma$ hyperons in $\Sigma p$ scattering experiments. It is designed for the new hyperon--proton scattering experiment conducted at J-PARC.
One critical reason behind the low statistics associated with hyperon--proton scattering experiments is the limited number of hyperon beams.
For example, in previous $\Sigma p$ scattering experiments conducted at KEK  \cite{Kondo:2000, Kanda:2005}, the number of tagged $\Sigma^{-}$ beams was only 1.8 $\times$ $10^{5}$; this limited number of beams resulted in poor statistical results regarding approximately 40 scattering events \cite{Kondo:2000}.
In the J-PARC experiment, we shall significantly improve the yield of the $\Sigma$ beam up to $8 \times 10^{7} $ using high-intensity $\pi^{\pm}$ beams of 20 M/spill in which the spill represents a beam cycle of 5.2 s with a beam extraction period of 2 s.
In such experimental conditions, CATCH is expected to operate stably under \color{black} a high singles count rate \color{black} of approximately 2 MHz and have a sufficiently fast time response to separate the scattering events from the accidental background.
In addition, a wider angular acceptance of the recoil proton is a  crucial aspect for maximizing the $\Sigma p$ scattering yield and the angular coverage.
Because the recoil proton is kinematically emitted to the forward angle in the laboratory frame, a longer structure toward the direction of the $\pi$ beam is desirable.
Conversely, to ensure that the magnetic spectrometer is placed as close as possible to the LH$_{2}$ target, CATCH must be reasonably short in the same direction to obtain a large acceptance of the scattered $K^{+}$, thereby maximizing the yield of the $\Sigma$ beam. 
To satisfy both requirements for the $\Sigma$ production and the $\Sigma p$ scattering, CATCH must be set up in a significantly compact manner  in the radial direction. 
We selected a detector combination involving a cylindrical scintillation fiber tracker and a calorimeter, as shown in Figure \ref{catch} to resolve both requirements.

The CFT , which reconstructs the trajectories of charged particles, is a main component of CATCH. 
It comprises eight cylindrical layers of scintillation fibers.
Each layer is placed coaxially from 49 mm to 84 mm in radius with a 5 mm interval in between the layers; the length of each layer in the direction of the beam is 400 mm.
The LH$_{2}$ target, stored in a cylindrical target vessel of diameter 40 mm and length 300 mm in the direction of the beam, is installed in the inner center of the CFT.
Owing to the flexibility of the arrangement of the fibers, the required compact setup is realized by placing eight fiber layers within a space of 35 mm in the radial direction.
The CFT has two different fiber configurations: the $\phi$ and $uv$ layers.
In the $\phi$ layers, the fibers are placed parallel to the direction of the beam around each layer's cylinder surface; these $\phi$ layers are used to measure the azimuthal angle ($\phi$).
Conversely, in the $uv$ layers, scintillation fibers are arranged in a spiral configuration along the surface of the cylinder.
\color{black}
The $u$- and $v$-layers have opposite tilt angles to the direction of the beam. 
\color{black}
The $uv$ layers are used to measure the zenith angle ($\theta$) of the track.
The trajectory of the charged particle obtained from the LH$_{2}$ target is then reconstructed three-dimensionally.
To identify the $\Sigma p$ scattering event, 
\color{black}
the angular resolution ($\sigma_{\theta}$) for the recoil proton measurement should satisfy $\sigma_\theta<2^{\circ} $.
\color{black}
In addition, the total amount of the fiber material must be as small as possible to ensure the detection of the low-energy protons down to approximately 20 MeV. 
Considering the aforementioned requirements, the diameter of the scintillation fibers was set to 0.75 mm. 
The necessary fast time response can be achieved by adopting the CFT. 
To distinguish triggered events from accidental coincidences of background events, the timing gate of the CFT must be narrower than 12 ns. 
Therefore, \color{black} a \color{black} time resolution ($\sigma_{time}$) \color{black} is required to be less than 2 ns \color{black} in $\sigma$. 

The kinetic energy of the recoil proton is measured using the calorimeter around the CFT.
The energy resolution must be higher than 3\% in the $\sigma$ for 80 MeV protons to enable the identification of $\Sigma p$ scattering. 
\color{black}
The crystal length should be 400 mm to cover the active region of the CFT in the beam direction. 
A high-density calorimeter is also preferable to stop protons up to 150 MeV.
\color{black}
In addition, a non-deliquescent crystal is essential for minimizing the material surrounding the crystal.
Considering the points presented above, we adopted a BGO-based scintillator as the calorimeter.
24 BGO crystals with a size of 30 (width) $\times$ 25 (thickness) $\times$ 400 (length) mm$^{3}$ were placed at the outer circumference of the CFT.
One concern involving the BGO calorimeter is its relatively longer decay time of approximately 300 ns.
The singles rate for one BGO crystal is estimated to be 40–-400 kHz for charged particles depending on the condition of the $\pi$ beam.
Under such a high singles rate, the pulse from the BGO calorimeter piles up with a specific probability.
Therefore, in the offline analysis, we adopted a waveform readout using a flash ADC to decompose the accumulated signals.

We placed a PiID  around the BGO calorimeter as the final component of CATCH. 
In the $\Sigma p$ scattering experiment, both the protons and $\pi^{\pm}$ particles from $\Sigma$ decay are detected using CATCH.
\color{black}
While the BGO calorimeter was designed to be sufficiently thick for stopping recoil protons of up to 150 MeV, many of swift pions penetrate the BGO calorimeter.
\color{black}
Therefore, the PiID provides additional information for identifying particles by determining whether charged particles are stopped at the BGO calorimeter.

\section{Structure and Development of CATCH}
In this section, we describe each component of CATCH in detail.

\subsection{Cylindrical Fiber Tracker (CFT)}

A scintillation fiber is a fiber-shaped plastic scintillator;  scintillation photons are transmitted to the edge of the fiber through a total reflection that occurs at the boundary between the inner core and the outer cladding with different refractive indexes.
\color{black}
Detecting the scintillation light at a fiber's edge face provides the hit information at the respective fiber's position.
Therefore, a multi-layered fiber tracker can facilitate the reconstruction of particle trajectories.
\color{black}
A scintillation fiber tracker comprises many scintillation fibers whose positions are well determined.
Therefore, the position information can be obtained by determining the correspondence between the fiber segment hit by the particle  and its fiber position.
Trajectories are reconstructed using multiple layers of arranged fibers.

\begin{figure}
\begin{center}
\includegraphics[width=75mm]{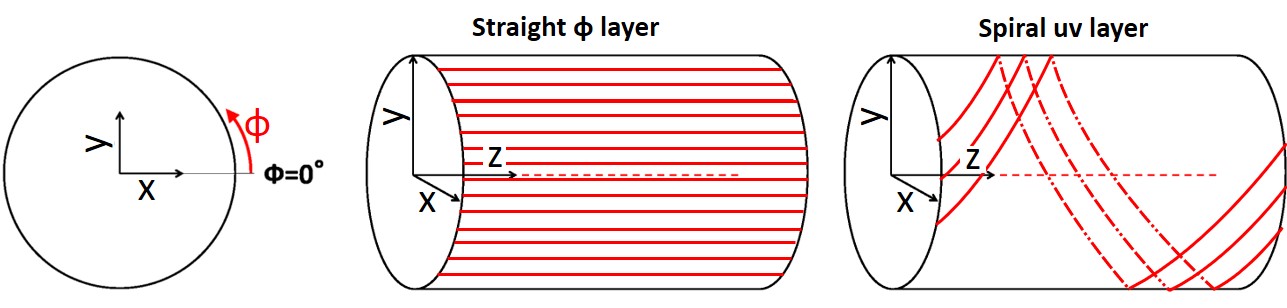}
\caption[Structure of fiber arrangement]{
Two types of fiber arrangements. The $\phi$ layer comprises fibers that are arranged parallel to the beam axis.
The $uv$ layer comprises fibers arranged along the surface of the cylinder that is centered on the beam axis.
}
\label{fiber_2type}
\end{center}
\end{figure}

The requisite characteristics of the scintillating fiber tracker employed in our proposed application are as follows:
First, the fiber can be arranged flexibly with regard to shape. 
Second, \color{black} its time response is sufficiently fast to achieve better time resolution than 2 ns\color{black}.
Finally, the energy deposited on the fiber can be measured.
In our application of the CFT, fibers are arranged alternately in a staggered manner along the cylindrical surface, and such fiber-based cylindrical layers are stacked coaxially in eight layers.
For layer configuration, we introduce two types of fiber arrangements, i.e., a straight $\phi$ layer and a spiral $uv$ layer in which fibers are arranged in parallel and spiral configurations based on the direction of the beam, as shown in Figure \ref{fiber_2type}. By combining the hit information regarding different fiber configurations, the trajectory of the charged particle can be reconstructed three-dimensionally. 

\subsubsection{Design of the CFT}

\begin{figure}
\begin{center}
\includegraphics[width=75mm]{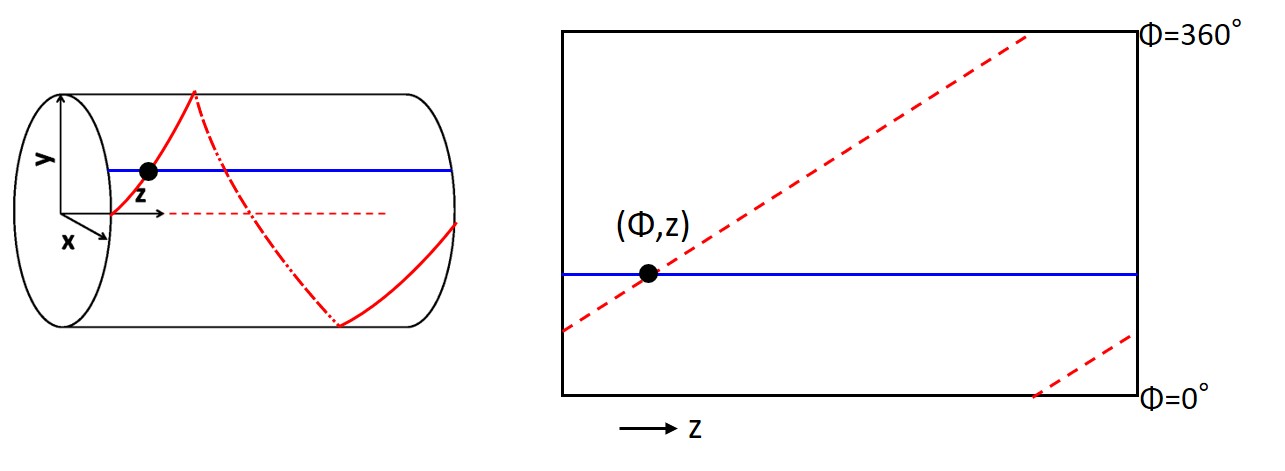}
\caption[Three-dimensional position of two fiber structures]{
Three-dimensional hit position of two fiber structures. 
The hit position in the direction of $\phi$ is obtained from the $\phi$ layers. The hit position in the $ z $ direction is then obtained by combining the hits of the $uv$ layers with the $\phi$ information.
}
\label{fiber_hit}
\end{center}
\end{figure}

In this subsection, the coordinate axes defined in Figure \ref{fiber_2type} are used to determine the cylindrical shape of the CFT.
The $xy$ plane is defined as the plane that is perpendicular to the central axis of the cylinder at the upstream edge of the active region of the CFT; the azimuthal angle along the circular direction in the $xy$ plane is referred to as $\phi$.
The $z$ axis is also defined as the central axis of the cylinder, and its origin is set to be the upstream edge.
Therefore, the center of the circle at the upstream edge is defined as the origin of the coordinate, i.e., ($x, y, z$) = (0, 0, 0).
The active region of the CFT, where the fibers are placed in the desired location, is designed to be 400 mm in the $z$ direction to cover the LH$_{2}$ target.
In the spiral $uv$ layers, fibers are arranged at a tilt angle to ensure that they rotate 360 degrees in the azimuthal angle $\phi$ from $z=0$ mm to $z=400$ mm in one turn, as shown in Figure \ref{fiber_2type}. 
Therefore there exists a correspondence between $\phi$ and $z$ for each fiber in the $uv$ layers.
Because the $\phi$ information regarding the $uv$ layer can be obtained through the tracking of four $\phi$ layers, the position information associated with the $z$ direction can be obtained, as demonstrated in Figure \ref{fiber_hit}.
We adopted a scintillation fiber of diameter 0.75 mm from the Kuraray  company (Kuraray SCSF-78M) by considering the angular resolution, the number of readout channels, and the ease of fabricating the CFT. 
The \color{black} radial distance \color{black} and the number of fibers for each installed layer are listed in Table \ref{fibers}.

\begin{table}
\begin{center}
\small
 \caption{\color{black} Radial distance \color{black} and number of fibers for each installed layer.}

\begin{tabular}{|c| |c|c|c|c|}
\hline
$\phi$ Layer & $\phi$1 & $\phi$2 & $\phi$3 & $\phi$4\\
\hline
\color{black} Radial distance \color{black}  [mm] & 54 & 64 & 74 & 84 \\
\hline
Number of fibers & 584 & 692 & 800 & 910 \\
\hline \hline

$uv$ Layer & $u$1 & $v$2 & $u$3 & $v$4\\
\hline
\color{black} Radial distance \color{black}  [mm] & 49 & 59 & 69 & 79 \\
\hline
Number of fibers & 426 & 472 & 510 & 538 \\
\hline
Tilt angle [degree] & 37.6 & 42.8 & 47.3 & 51.1 \\
\hline
\end{tabular}

\end{center}
\label{fibers}
\end{table}

\subsubsection{Fabrication of the CFT}

\begin{figure}
\begin{center}
\includegraphics[width=75mm]{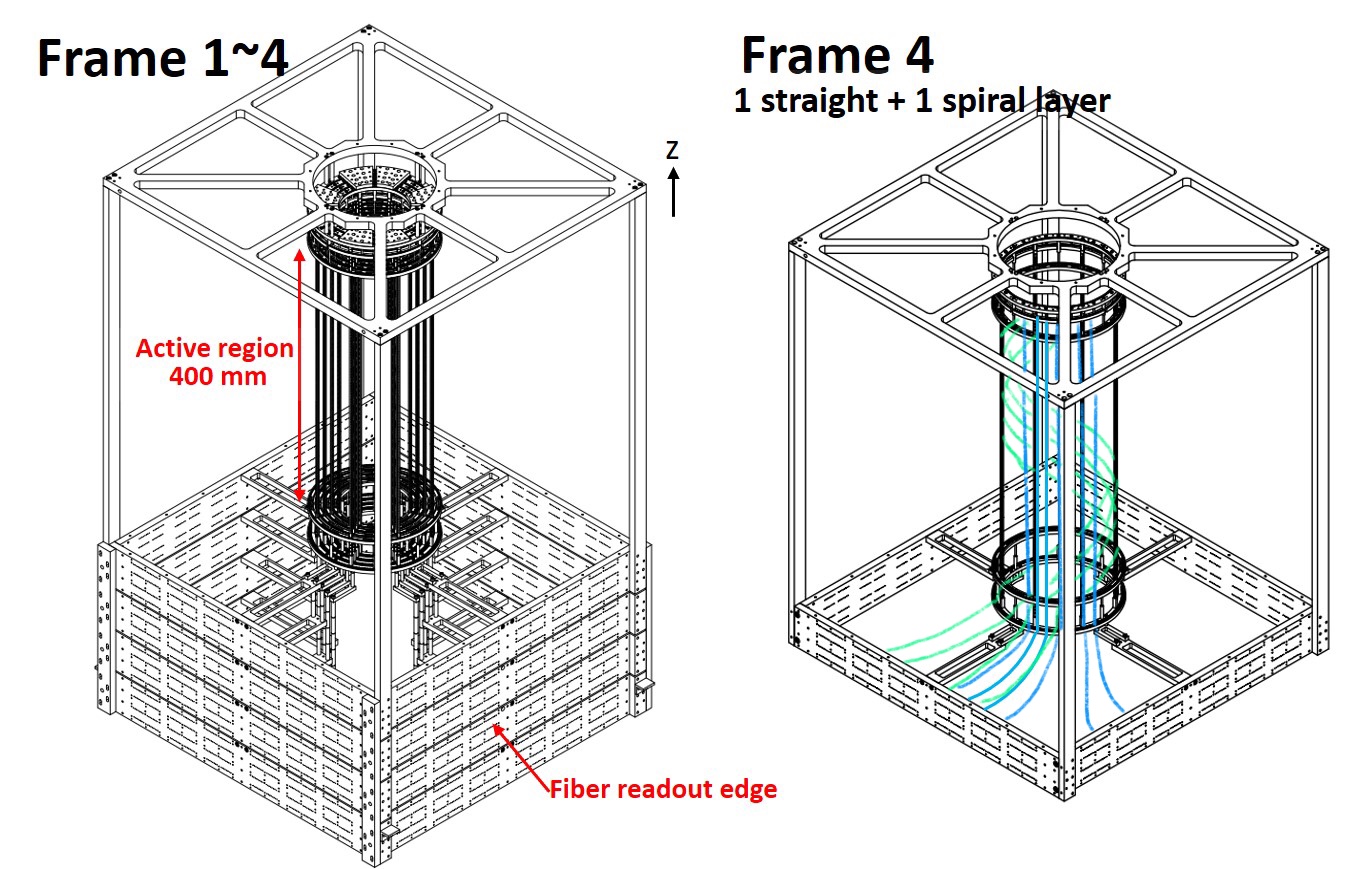}
\caption[CFT straight layer structure]{
Design of the CFT frame. The CFT comprises $4$ frames, including one $\phi$ layer and one $u$ or $v$ layer for each frame to realize a compact layer structure. 
}
\label{CFTframe}
\end{center}
\end{figure}

\color{black}
From the fabrication perspective, developing such cylindrical structures is very  difficult because of the narrow spacing between layers.
\color{black}
To address this difficulty, we fabricated four components containing one $u$ or $v$ layer and one $\phi$ layer, separately.
In each component, fibers with two different layer configurations were attached to the detector frame, as shown in the figure on the right side of Figure \ref{CFTframe}.

\begin{figure}
\begin{center}
\includegraphics[width=75mm]{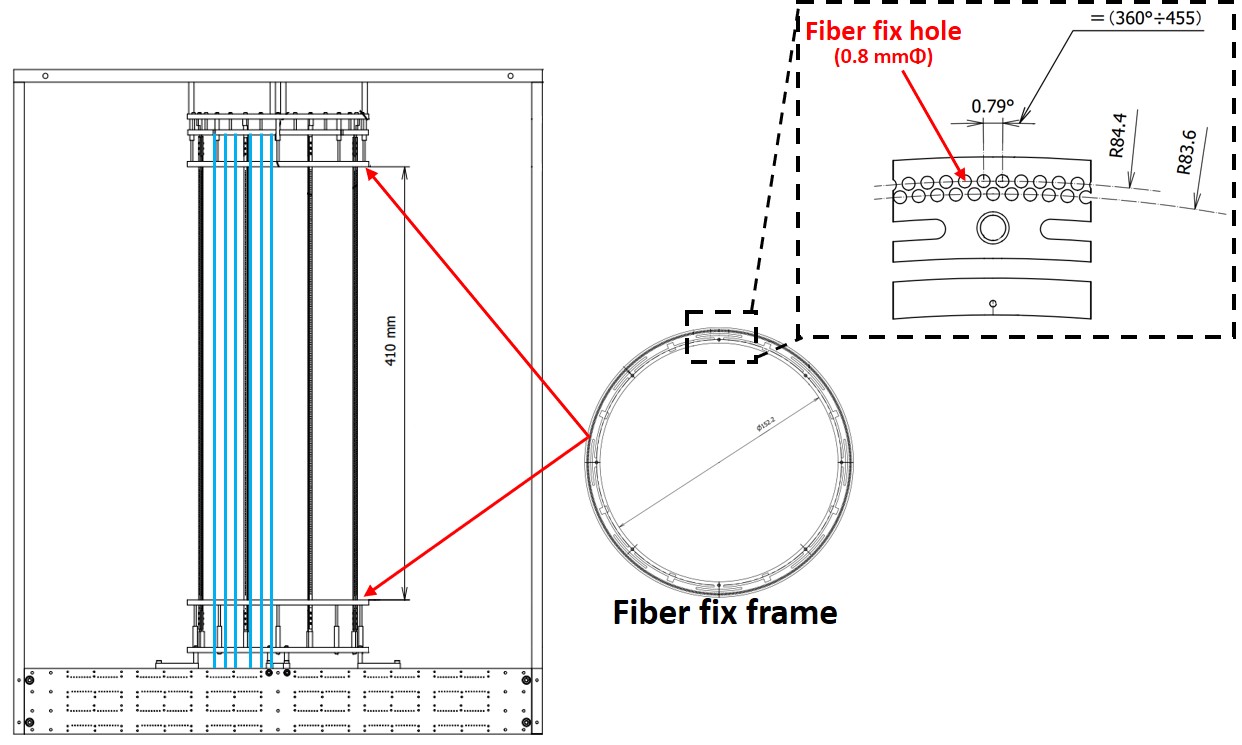}
\caption[CFT structure with four straight layers]{
Fiber-fixing frames with holes for fiber to fiber attachment.  These were placed at both ends of the active region. 
}
\label{phi4_frame}
\end{center}
\end{figure}

\begin{figure}
\begin{center}
\includegraphics[width=70mm]{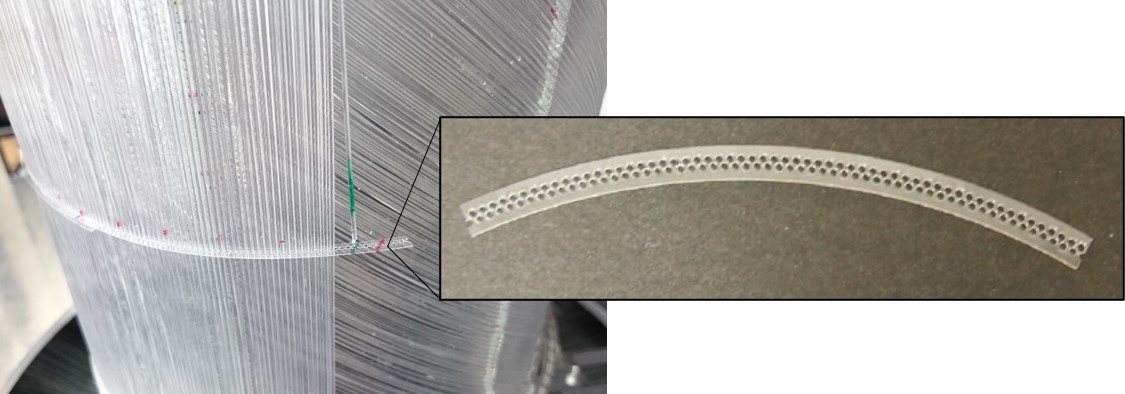}
\caption[Mylar sheet]{
Photo of a thin Mylar sheet of 0.1 mm thickness with holes corresponding to the fiber position.
}
\label{Mylar}
\end{center}
\end{figure}

\begin{figure}
\begin{center}
\includegraphics[width=75mm]{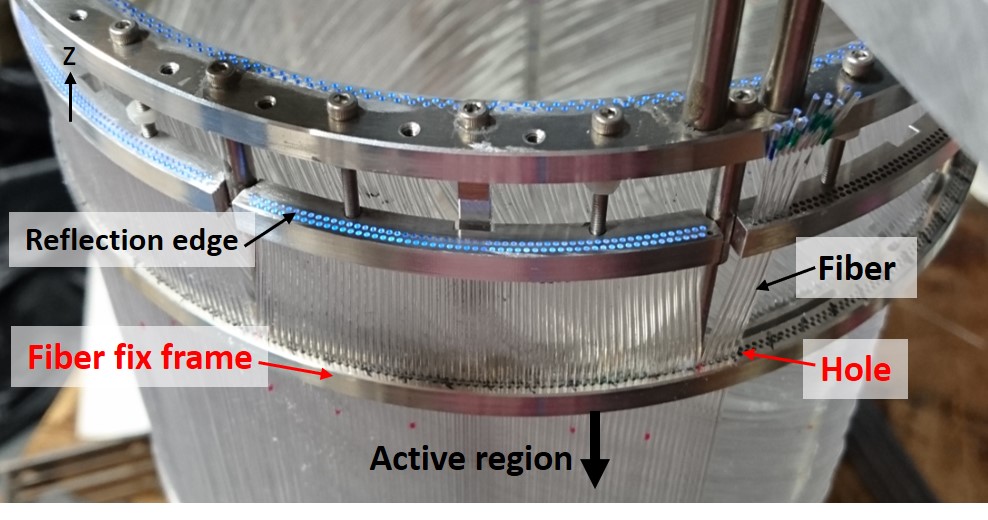}
\caption[Picture of straight layer frame (reflection side)]{
Fiber-fixing frame at the downstream to determine fiber positioning. 
The fiber tension can be adjusted by adjusting the screws connected to the tension-adjustment frame (Reflection edge).
}
\label{framepic}
\end{center}
\end{figure}

The actual fabrication procedure for the straight $\phi$ layer is as follows.
In the $\phi$ layer, the fibers must maintain their position over the entire active area of 400 mm in the $z$ direction.
The misalignment of the fibers results in lower efficiency and worsens the angular resolution compared to that of the ideal conditions.
One technique for maintaining the position of fibers involves ensuring that the neighboring fibers are attached to one other using an adhesive to form a cylindrical shape. 
However, this method increases the need for an adhesive material.
This is not suitable for our proposed application, because the detection efficiency for  lower-energy protons decreases due to the additional material.
To address the difficulty of fiber alignment, we prepared frames and position sheets to determine the fiber position.
The fiber positions at both edges of the active region were determined using the fiber-fixing frame in which holes with a diameter of 0.80 mm were drilled along the fiber position, as shown in the enlarged figure of Figure \ref{phi4_frame}. 
As can be seen in the figure, fibers are arranged in a staggered structure to minimize the ineffective area owing to the gap between the fibers.
To maintain the position over the active region, thin Mylar sheets of 0.1 mm in thickness, with holes corresponding to the fiber position, were inserted at intervals of 50 mm, as shown in Figure \ref{Mylar}.
If the fibers are stretched correctly with the appropriate tension, they are placed in the desired position.
For this purpose, we also added a function for adjusting the fiber tensions to the frame structure, as shown in Figure \ref{framepic}. 
The fiber-fixing frame at the downstream ($z=400$ mm), as shown in Figure \ref{framepic}, is used only to determine the fiber position; the fibers are not attached to this hole.
Alternatively, the fibers are attached to the tension-adjustment frame, as shown in Figure \ref{framepic}, whereas fibers on the other side ($z=0$ mm) are attached to the fiber-fixing frame.
Optical cement was used as the adhesive material for attaching fibers to frames. Fiber tension can be adjusted by adjusting the screws connected to the tension-adjustment frame. 


\begin{figure}
\begin{center}
\includegraphics[width=80mm]{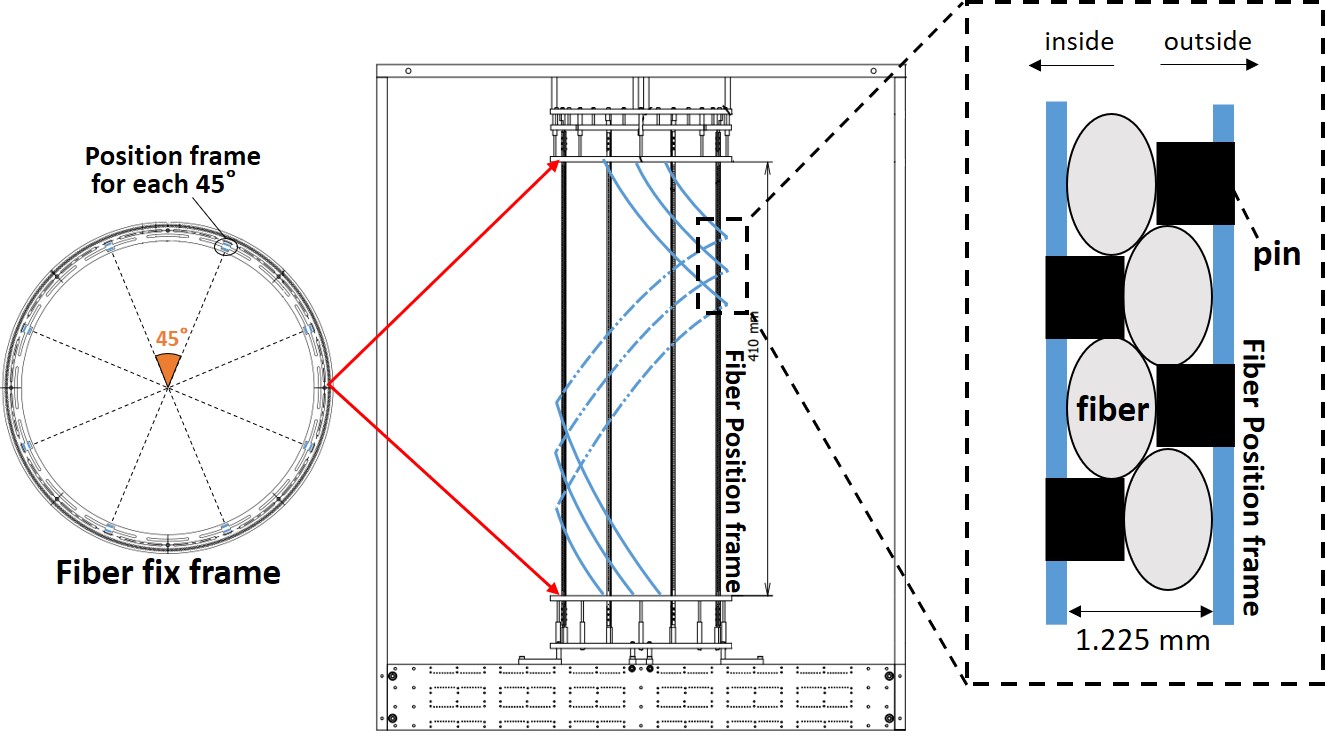}
\caption[Fabrication method of the CFT spiral layer]{
 Fiber-positioning pillars at eight circular positions set at 45$^{\circ}$ intervals. The $ z $ position of each fiber at the pillar location was determined. The pillar comprises two acrylic bars and position pins. 
}
\label{CFTframeUV}
\end{center}
\end{figure}

\begin{figure}
\begin{center}
\includegraphics[width=60mm]{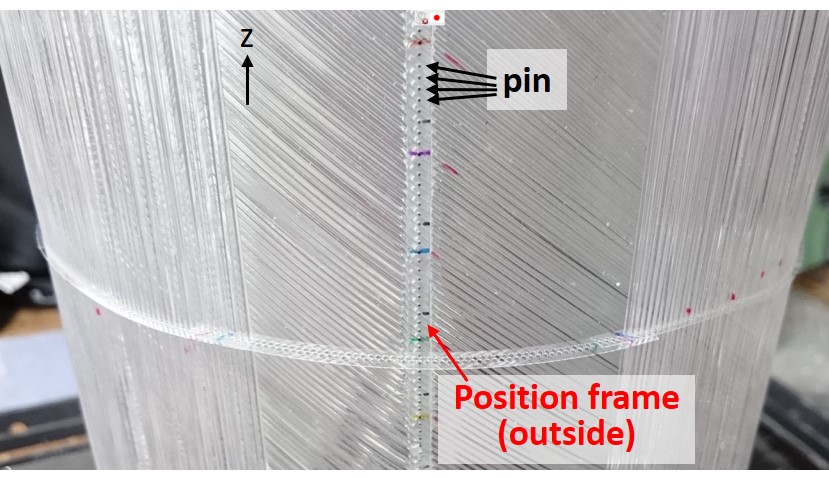}
\includegraphics[width=60mm]{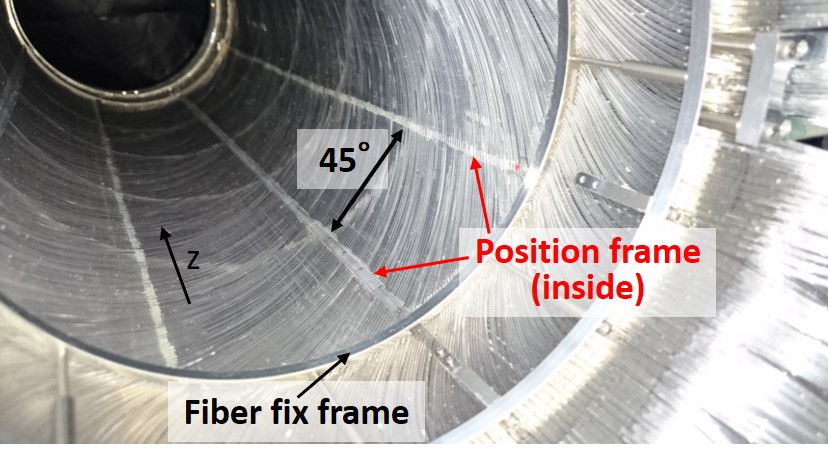}
\caption[Picture of the CFT spiral layer during fabrication]{
Fabricated spiral layers observed from the outside (top) and the inside (bottom), respectively.
}
\label{outside}
\end{center}
\end{figure}

\color{black}The \color{black} fabrication procedure for the spiral $uv$ layer is as follows.
The fibers in the $uv$ layer have to be arranged in a specific tilt angle along the cylindrical surface.
To realize the fabrication of such a challenging structure, we decided to introduce fiber-positioning pillars at eight circular positions set at 45$^{\circ}$ intervals, as shown in Figure \ref{CFTframeUV}. 
The pillar is made of an acrylic bar of 1 mm thickness, and it has position pins of 400 $\mu$m diameter to determine the fiber position, as shown in the enlarged figure of Figure \ref{CFTframeUV}.
The fibers are sandwiched by pairs of positioning pillars with position pins; they have a zigzag configuration to minimize the ineffective area.
Considering the thickness of the material, it is better to remove these pillars as much as possible.
Positioning pillars were used throughout the fabrication process.
After placing all the fibers, a thin layer of epoxy glue was pasted to a limitted region along the  pillars to maintain the spiral shape.
Finally, half of the positioning pillars were removed, thereby leaving four of them at 90$^{\circ}$ intervals.
Figure \ref{outside} shows the fabricated spiral layers in which fibers are arranged in the desired spiral configuration.

\begin{figure}
\begin{center}
\includegraphics[width=75mm]{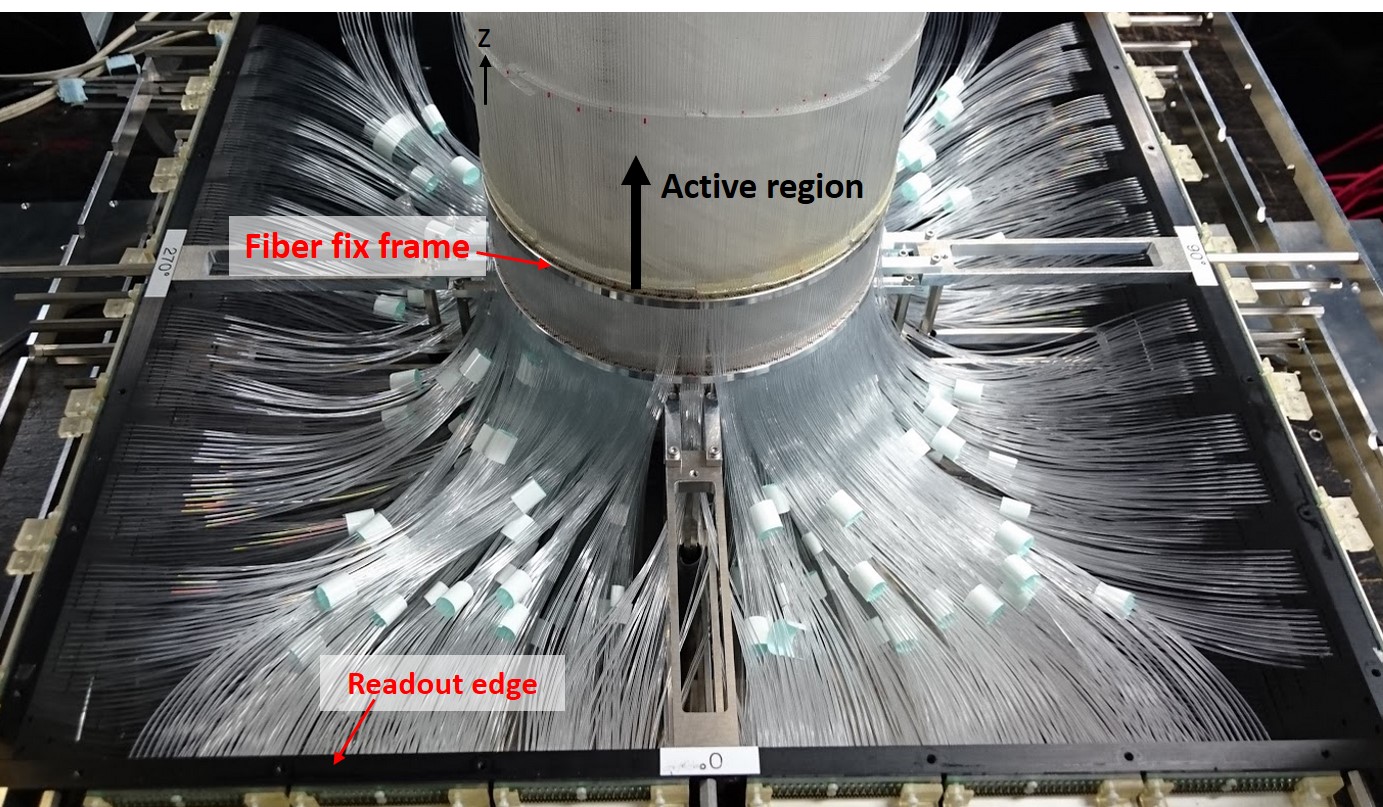}
\caption[Picture of the straight layer frame (readout side)]{
Photo of the fibers attached to the readout frame after passing through the holes of the fixed frame.
}
\label{pic_readout}
\end{center}
\end{figure}

In one component, one spiral $u$ or $v$ layer and one straight $\phi$ layer are placed at the inner and outer sides in the common detector frame, respectively.
These fiber edges at the downstream side ($z > 400$ mm) are attached to the detector frame, as shown in Figure \ref{framepic}.
To increase the light yield by reflecting the photon transmitted to this edge, an enhanced specular reflector (ESR) sheet was attached to this surface as the reflector.
The other edge ($z < 0$ mm) of the fiber is the readout side of the photon.
This readout edge of the fiber is guided to a hole in the readout frame, as shown in Figure \ref{pic_readout}.
The fiber position at the readout frame corresponds to the position of each photon sensor, which will be described in Section \ref{VME-EASIROC}.
To obtain a flat surface for improved light transmission, the readout surface was polished mechanically through diamond polishing.

\begin{figure}
\begin{center}
\includegraphics[width=75mm, angle=90]{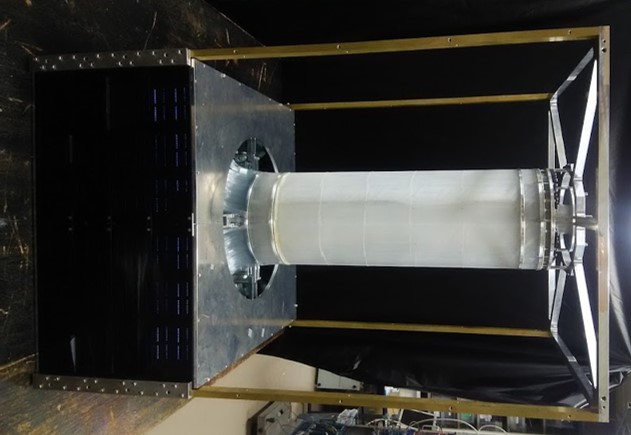}
\caption[Picture of the CFT]{
Photo of the CFT after combining all the layers.
}
\label{CFTpic}
\end{center}
\end{figure}

These four CFT components were fabricated separately.
Finally, the most inner component was inserted into the neighboring outer component, and these two components were attached to one another.
This process was repeated until all the four components were connected, as shown in Figure \ref{CFTpic}.

\subsubsection{MPPC readout system using the EASIROC board} \label{VME-EASIROC}
\begin{figure}
\begin{center}
\includegraphics[width=50mm]{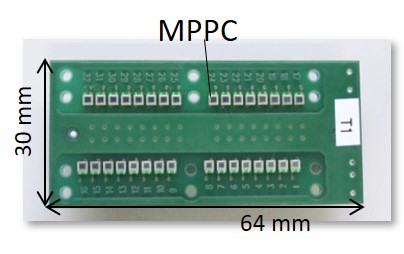}
\includegraphics[width=70mm]{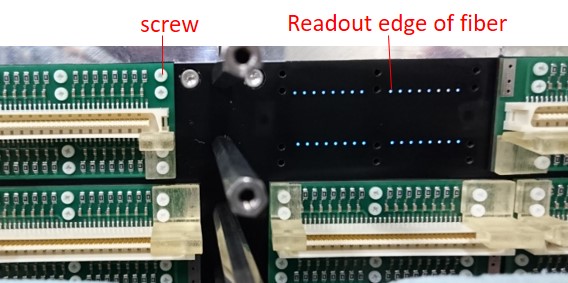}
\caption[DAQ system]{
Readout circuit board on which 32 MPPCs are mounted.
Bottom picture shows contact between the fibers and the MPPCs. 
}
\label{MPPC}
\end{center}
\end{figure}

\ \\
The scintillation lights in each CFT fiber are read out one after the other using an multi-pixel photon counter (MPPC)  (Hamamatsu S10362-11-050P) whose active area is 1$\times$1 mm$^{2}$ with 400 pixels of the avalanche photo diode (APD).
Because approximately 5,000 fibers are used for the CFT, the same amount of MPPCs should be read out.
We developed a printed circuit board (MPPC PCB) on which 32 MPPCs are mounted in 2.54 mm pitch, as shown in Figure \ref{MPPC}.
To ensure satisfactory contact between the fibers and the MPPCs, the MPPCs are placed with a precision of more  than 100 $\mu$m on the MPPC PCB.
The fibers are attached to the holes on the readout frame, where the positions of the holes correspond to the MPPC position on the PCB.
Using screws to attach the MPPC PCB to the readout frame, the contact between the fibers and the MPPCs is ensured.
In total, 157 PCBs are used for the readout of the fibers in the CFT.

\begin{figure*}
\begin{center}
\includegraphics[width=120mm]{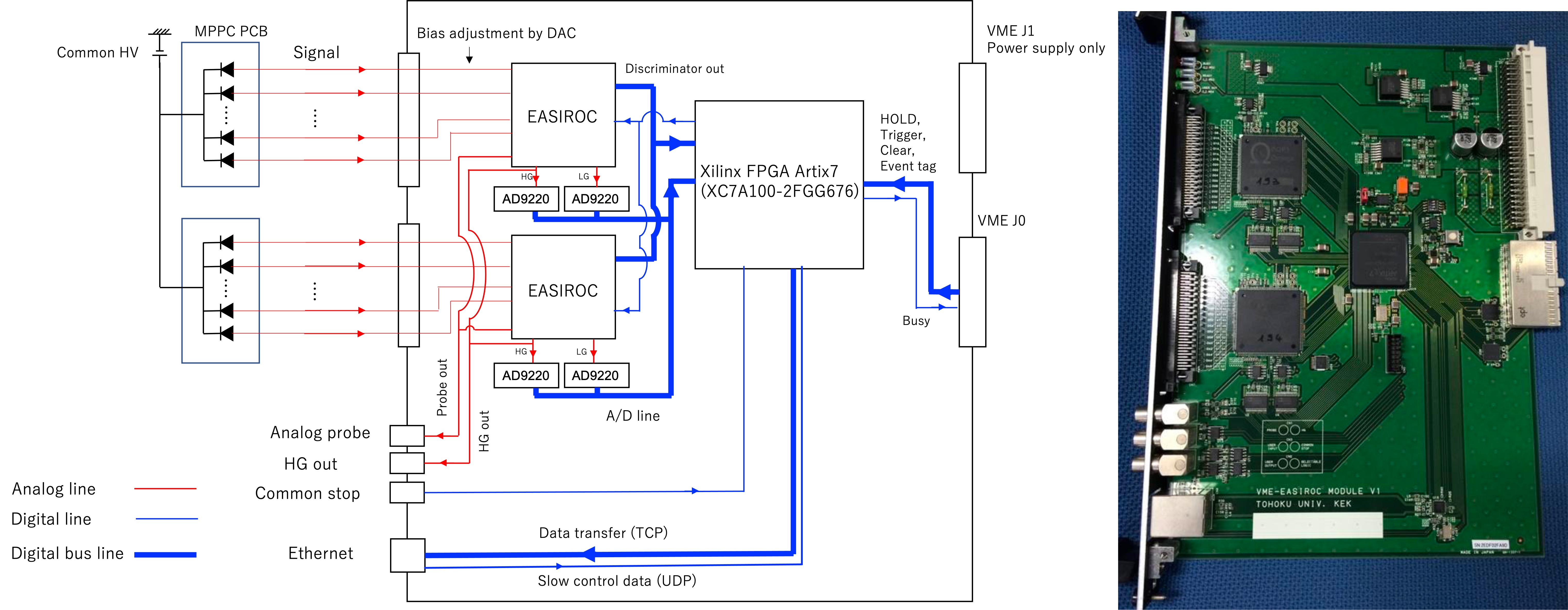}
\caption[Block diagram of VME-EASIROC]{
Block diagram and photo of VME-EASIROC.
Two EASIROC chips are mounted on the board, and these chips are controlled using the Xilinx FPGA Artix7.
The red and blue lines show the analog and digital lines, respectively, and the bold blue line shows the digital bus line. 
Trigger signals are input from the front connector and the VME J0 connector in the backplane.
The communication between the PC and VME-EASIROC is performed via Ethernet with a throughput of 100 Mbps. 
}
\label{VME_EASIROC}
\end{center}
\end{figure*}

To handle such a large amount of MPPCs, we developed a dedicated readout board using an application specific integrated circuit (ASIC)  of the extended analogue silicon photomultipliers integrated readout chip (EASIROC) \cite{EASIROC}, which was developed by Omega/IN2P3 to handle the multi-pixelated photon detector (PPD).
MPPC is a PPD produced by Hamamatsu photonics.
The EASIROC has preamplifiers, shapers, and a discriminator for each of the 32 channel inputs.
We developed a readout board designed for the CFT using the EASIROC, called VME-EASIROC.
The block diagram and photo of VME-EASIROC are shown in Figure \ref{VME_EASIROC}.
In the VME-EASIROC board, 64 channels of MPPCs can be read out by mounting two EASIROC chips on the board.
The control of the EASIROC chips and the processing of the digital signals are handled by the Xilinx FPGA Artix7 (XC7A100-2FGG676) mounted on the VME-EASIROC board.
A common voltage from an external power supply is applied to the MPPC cathodes, whereas the anode voltage can be adjusted using the digital-to-analog converter (DAC)  in the EASIROC chip with a precision of approximately 20 mV per channel.
In the EASIROC chip, two analog lines with different amplification gain settings exist; both analog lines are followed by a charge preamplifier, a slow shaper, and a switched capacitor for a charge measurement.
The analog line with a high gain setting (HG line) is optimized to measure small amounts of photo-electrons, whereas that with a low gain setting (LG line) is suitable for measuring large amounts of photo-electrons.
The shaping time of the slow shaper is adjustable from 25 ns to 175 ns.
By inputting a trigger signal (HOLD), the pulse height of the slow shaper is maintained in the switched capacitor.
In our application, we used the slowest shaping time of 175 ns to ensure that the slow shaper pulse became the peak at the timing of the HOLD trigger.
The 32 channels of the switched capacitor for each analog line make up an analog multiplexer and two analog multiplexers for the HG and LG lines, respectively, which are connected to external ADC chips (AD9220).
The A/D converted 12-bit binary data are sent to FPGA , and thus, the charge information of both the HG and LG lines for all channels are obtained simultaneously.
To suppress the huge amount of data, the pedestal suppression function is also implemented in FPGA.
For the timing measurement, the HG line is also followed by a fast shaper with a shaping time of 15 ns.
The fast shaper signal is distinguished through the update type discriminator with a common threshold for all 32 channels in one EASIROC chip; the threshold is set by the register value of the threshold DAC.
The logic signals from the discriminators are fed into FPGA in parallel to measure its timing and time over threshold (TOT) using FPGA-based multi-hit TDC  (MHTDC).
This MHTDC measures the timing of leading and trailing edges of signals with a timing precision of one ns.
Data communication between the PC and the VME-EASIROC board is performed via Ethernet with a throughput of 100 Mbps using a function of a hardware-based TCP processor (SiTCP) implemented in FPGA \cite{Uchida:2008}.
The detector data, including the charge and time measurements are transmitted through SiTCP by generating Ethernet frames from the detector data.
Conversely, the slow control for setting the register of the EASIROC chip is performed through a user datagram protocol (UDP) in the SiTCP function.

In total, 79 VME-EASIROC boards are used for the CFT readout.

\subsection{BGO calorimeter}

\begin{figure}
\begin{center}
\includegraphics[width=70mm]{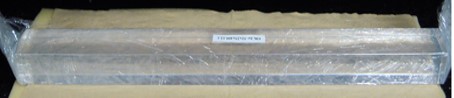}
\caption[Picture of a BGO crystal]{
Photo of the BGO crystal, whose size is $30 \times 25 \times 400$ mm$^3$, for the scintillation calorimeter.
}
\label{bgo}
\end{center}
\end{figure}

\begin{figure}
\centering
\includegraphics[width=80mm]{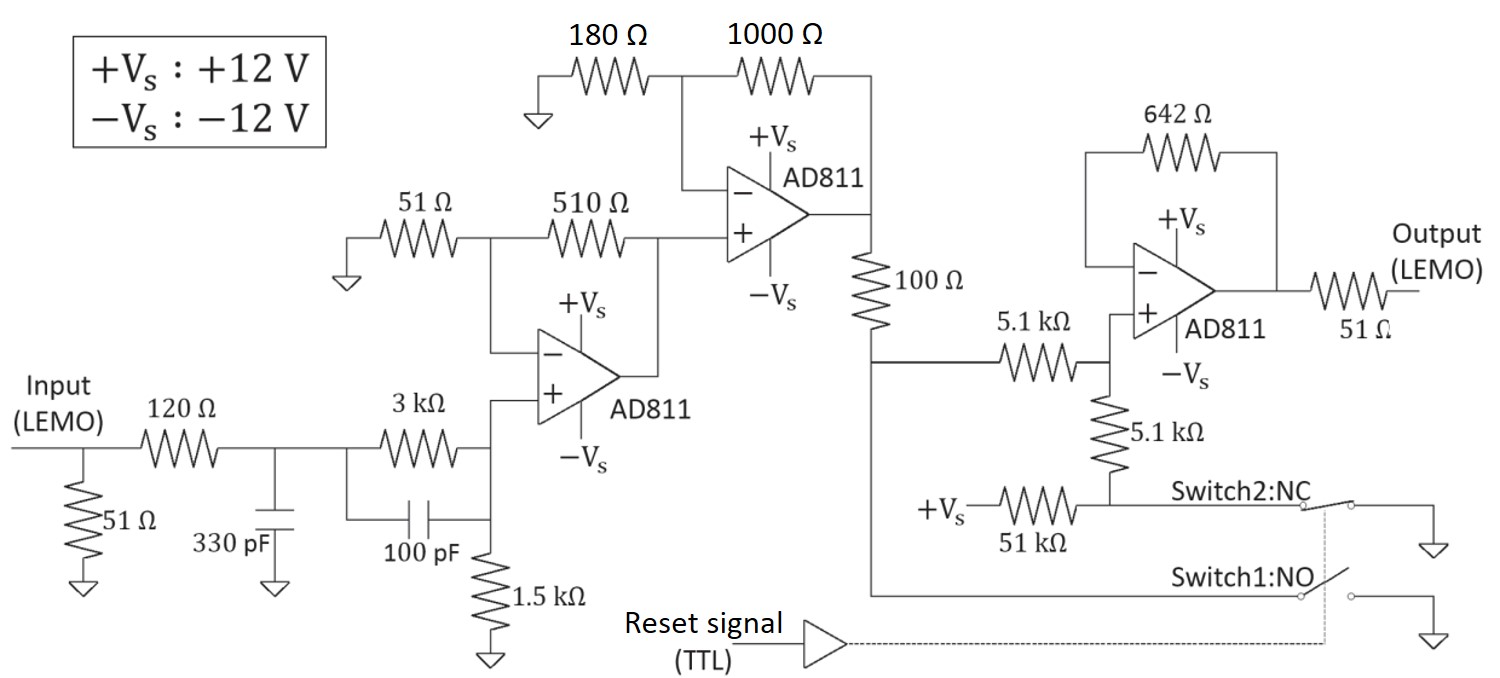}
\caption[BGO circuit diagram]{
Circuit diagram of the developed shaping amplifier, which has the shaping circuit of the integral filter and an amplifier circuit to compensate for the lower gain of PMT under the lower voltage operation.
}
\label{ikd}
\end{figure}

\begin{figure}
\centering
\includegraphics[width=70mm]{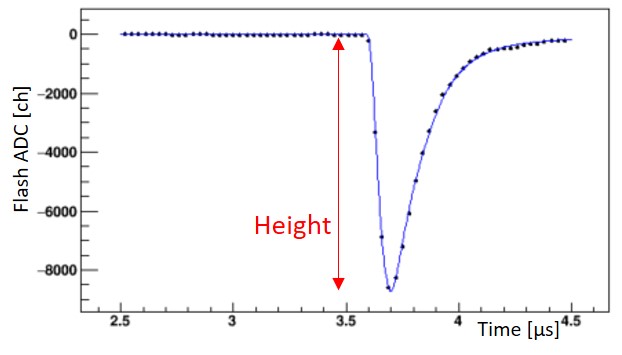}
\caption[Waveform analysis for BGO calorimeters]{
Typical waveform of the BGO calorimeter after the shaping amplifier circuit. 
The black points show the flash ADC data with a sampling rate of 33 MHz.
The blue line shows the reconstructed pulse shape by fitting the data point with a template waveform. 
}
\label{bgo_wave}
\end{figure}

The BGO calorimeter comprises 24 BGO crystals whose size is 30 (width) $\times$ 25 (thickness) $\times$ 400 (length) mm$^{3}$; it is placed outside of the CFT.
Figure \ref{bgo} shows the photograph of the BGO crystal used as the calorimeter.
Photomultiplier tubes (PMT) (Hamamatsu H11934-100) are used as the readout sensors.
The BGO calorimeter must have an energy resolution higher than 3\% ($\sigma$) for 80 MeV protons under the expected high singles rate of 40 to 400 kHz in the $\Sigma p$ scattering experiment.
Under such a singles rate, the pulse from the BGO calorimeter frequently accumulates owing to the relatively long decay time of 300 ns.
To decompose the accumulation events, we selected the waveform readout using a flash ADC.
If we consider the waveform data of the raw BGO signal, a high sampling rate is necessary to detect the fast leading edge of the signal.
Because the BGO signal is significantly long ($\sim$1 $\mu$s ) in total, the data size becomes very large.
To reduce the data size, the BGO signal is filtered using an integral circuit and the shaped signal is sampled at a low sampling rate of a few 10 ns.
Because the sampling points are coarse, the original pulse shape is reproduced by fitting the sampled waveform data with a template function to maintain the energy resolution.
One more concern under the high singles rate is the stability of PMT  under the high current at the dynode.
If the rated voltage of 900 V is applied for PMT under such high rate conditions, the gain of PMT fluctuates significantly owing to the voltage drop at the final dynode in PMT as a result of the large dynode current.
The instability of the PMT gain results in the deterioration of the energy resolution of the BGO calorimeter.
To suppress the instability, PMT should be operated at \color{black} a \color{black} lower HV of approximately 600 V.
We developed a shaping amplifier circuit, which has the shaping circuit of the integral filter, and an amplifier circuit to compensate for the lower gain of PMT under the lower voltage operation.
Figure \ref{ikd} shows the circuit diagram of the developed shaping amplifier.
The first component is the integral circuit whose time constant is set to 39.6 ns followed by the pole zero cancellation circuit to shorten the falling time.
The second component involves the two stages of the non-inverting amplifier circuit, where the operational amplifier AD811 is used.
The typical amplification factors are 11 and 6.6 for \color{black} each stage, \color{black} respectively.
The final stage is a signal reset function, which forces the output signal to be almost grand level by inputting a reset signal in the Transistor-Transistor Logic (TTL)  level.
This function is prepared for data suppression by combining the pedestal suppression function in a flash ADC module (CAEN V1724). However, it was not used in the experiment.
Figure \ref{bgo_wave} shows the waveform of the output signal of the shaping amplifier circuit. 
The rising time is approximately 100 ns, and the falling time from the peak to the 10\% level of the pulse peak is 370 ns.
The minimum sampling points at the leading region are three points, which are used to obtain the required energy resolution by reconstructing the original pulse shape and by fitting it with the template function of the waveform.
Therefore, the sampling frequency of the flash ADC was adopted to be 33 MHz.
Through this readout method, the energy resolution was 1.3\% \color{black} at 80 MeV with the proton intensity of up to 550 kHz in the test experiment.\color{black}

\subsection{PiID counter}

\begin{figure}
\centering
\includegraphics[width=80mm]{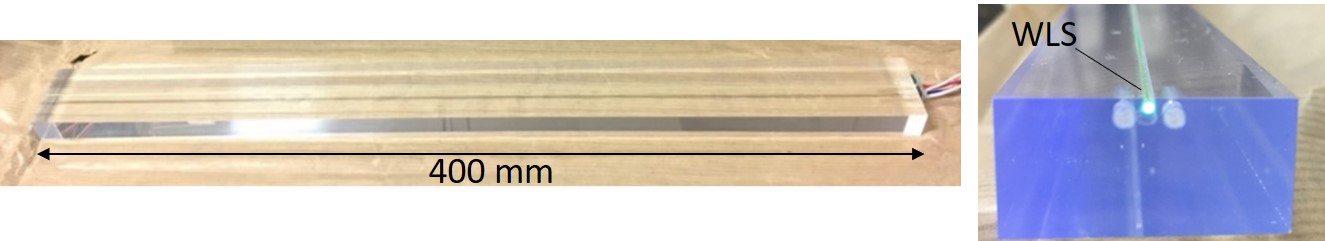}
\caption[PiID]{
Photo of the PiID counter.
It comprises a scintillator of 30 (width) $\times$ 15 (thickness) $\times$ 400 (length) mm$^{3}$ and a wavelength shifting fiber 1.0 mm in diameter mounted in a groove dug in the scintillator. 
}
\label{PiID}
\end{figure}

For satisfactory particle identification, we also install a PiID counter, which is a plastic scintillator located at the outside of the BGO calorimeter.
Basically, the particle identification is performed using the partial and total energy deposit information ($\Delta E$-$E$) in the CFT fibers ($\Delta E$) and the BGO ($E$) calorimeter. Because almost all $\pi$ particles penetrate the BGO calorimeter, the hit information of the PiID counter helps in the particle differentiation  between $\pi$ and the proton. We adopted plastic scintillators with a wavelength shifting fiber for a MPPC readout. The PiID counter comprises 34 plastic scintillators whose size is 30 (width) $\times$ 15 (thickness) $\times$ 400 (length) mm$^{3}$. A wavelength shifting fiber (WLS) of 1.0 mm diameter from the Kuraray company (Kuraray Y-11(200)M) is mounted in a groove dug in the scintillator, as shown in Figure \ref{PiID}. 
Photons of the WLS are read out using an MPPC (Hamamatsu S10362-11-100P) attached to the scintillator with screws.
The VME-EASIROC board is also used for the readout of the PiID counter.

\section{Commissioning of the CFT using cosmic rays}
The CFT was fabricated by placing fibers one after the other, as described in Section 3. 
We focused on the fiber’s position and its tension to maintain the desired position. 
However, there was a limit on the accurate placement of fibers.
In the experimental setup, where the CFT is placed horizontally, the fibers are also deflected under its  weight.
Therefore, the actual fiber position should be estimated in other ways.
For this purpose, we performed the CFT operation with cosmic ray measurement to collect calibration data for the CFT fiber positions.
In this section, we focus on the correction of the fiber position with the cosmic ray data.

\subsection{Cosmic ray measurement using the CFT}
\begin{figure}
\centering
\includegraphics[width=70mm]{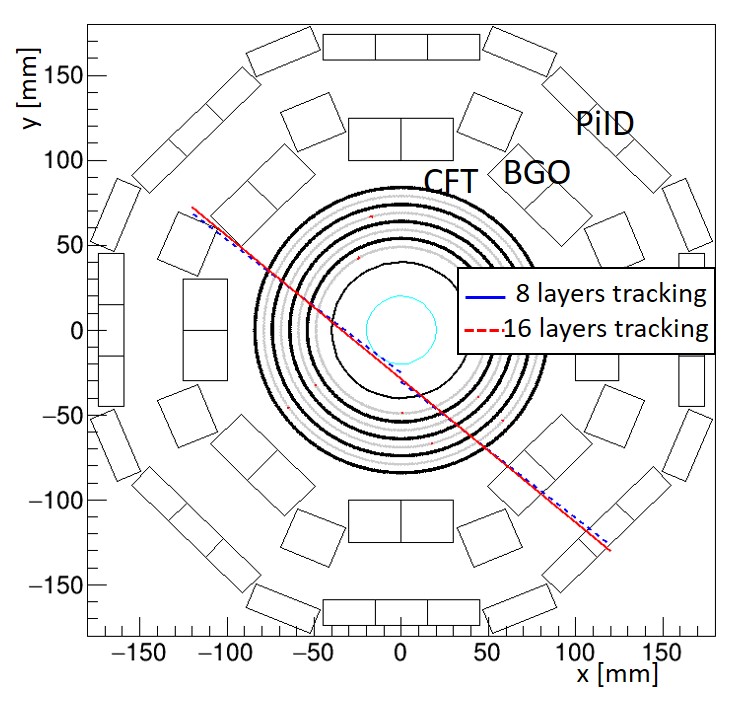}
\includegraphics[width=70mm]{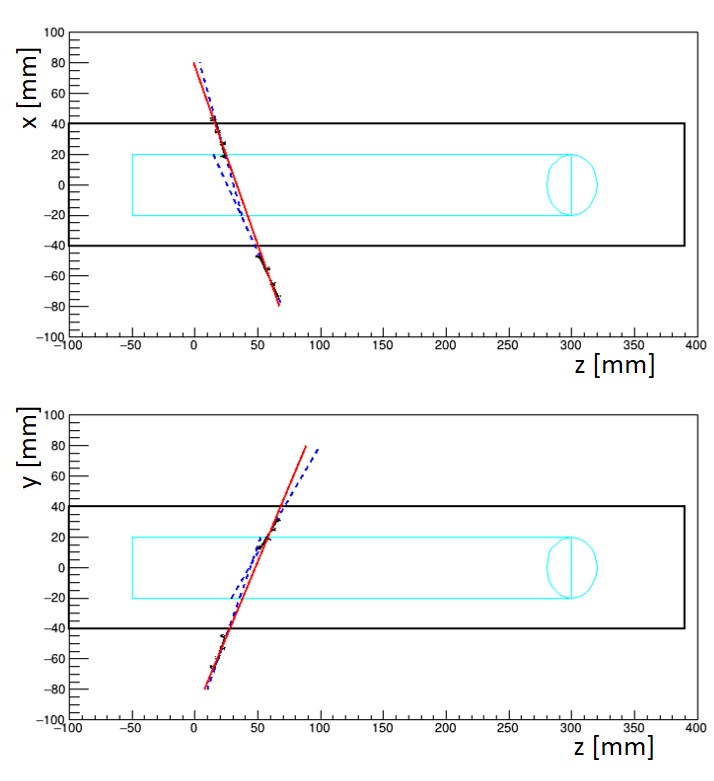}
\caption[Event Display]{
Example of the cosmic ray event detected using CATCH. The tracks of the cosmic ray reconstructed from the 16 layers and eight layers are also shown using red solid lines and blue dashed lines, respectively.
}
\label{EvDisp}
\end{figure}

\begin{figure}
\centering
\includegraphics[width=80mm]{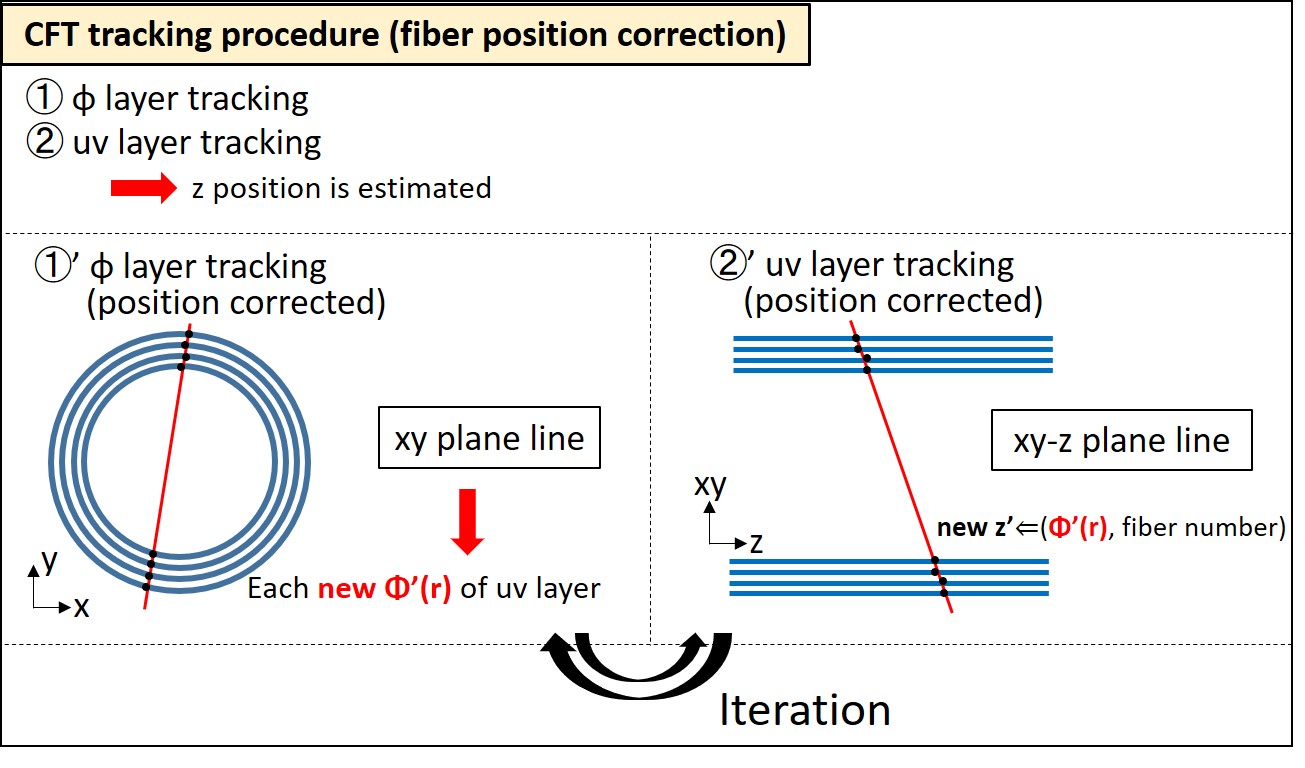}
\caption[CFT procedure with fiber position correction]{
Tracking procedure of the CFT with two stages. 
To reproduce the actual fiber position, the position correction was iterated because the tracking of the $\phi$ and $uv$ layers is related.
}
\label{tracking_c}
\end{figure}

We accumulated cosmic ray data using the full setup of CATCH for approximately 10 days.
In the entire operation of the CFT, 22 segments of fibers were missing owing to problems in the MPPC and the scintillation fiber itself, out of 4,932 fibers.
These missing channels affected the detection efficiency of the CFT; the efficiency will be discussed later.
For data acquisition, the coincidence between the two inner fiber layers ($\phi1$ and $\phi2$) was used as the trigger.
 The ADC and TDC data of the CFT were read out using the VME-EASIROC board for each fiber to reconstruct the trajectories of the cosmic rays.
To identify the hits in the fibers, the hit timing obtained from the TDC was required to be within $\pm$20 ns from the triggered timing, as discussed in the following analysis.
Figure \ref{EvDisp} shows an example of the cosmic ray event detected using CATCH. 
As can be seen in this figure, the track of the cosmic ray can be reconstructed from hits in all the 16 layers in combination with the top eight layers and the bottom eight layers.
Therefore the actual trajectory can be estimated with improved accuracy even if the fiber position used in the tracking is different from the actual position resulting from the misalignment.
To obtain trajectories using the CFT, two stages of tracking are performed, as shown in Figure \ref{tracking_c}. First, a tracking in the $xy$ plane is performed using the $\phi$ layers. By fitting the hit points of each $\phi$ layer with a straight line, the trajectory in the $xy$ plane can be obtained. Next, straight tracking in the $zx$ or $zy$ plane is performed using the hit information in the $uv$ layer. The $z$ positions in the $uv$ layers can be estimated from the one-to-one correspondence between the $z$ position and the azimuthal angle ($\phi$), which is obtained from the intersection between the straight track in the $xy$ plane and the circle with the radius of each $uv$ layer.
The trajectory in the $zx$ or $zy$ plane is then obtained by fitting the hit points of each spiral layer. Therefore, the trajectory can be reconstructed three-dimensionally using the CFT. 
For position calibration, the deviation between the predicted position from the tracking and the used fiber position was verified for all the fibers one after the other; the fiber position was corrected to ensure that the deviation was small.
We iterated the tracking using the corrected fiber position and the correction of the fiber position after the tracking until the deviation was negligible.

\begin{figure}
\centering
\includegraphics[width=50mm]{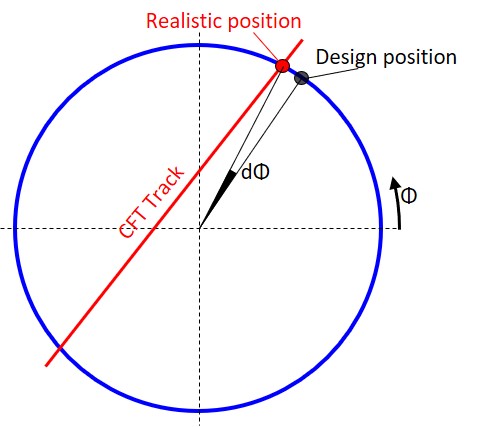}
\caption[Definition of d$\phi$]{
Definition of the deviation in $\phi$ ($d\phi$) in the $xy$ plane. $d\phi $ represents the difference in the azimuthal angle $\phi$ between the desired position and the estimated position from the tracking process. 
}
\label{dphi_zu}
\end{figure}

\begin{figure}
\centering
\includegraphics[width=70mm]{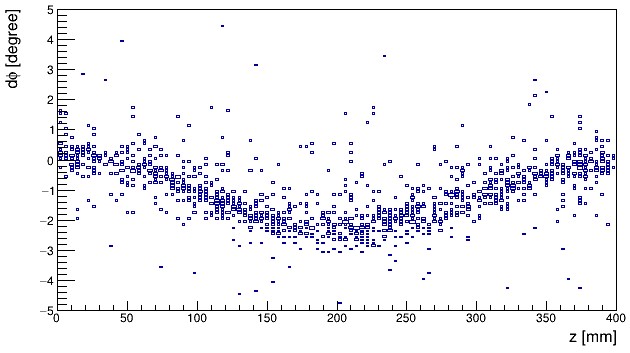}
\includegraphics[width=70mm]{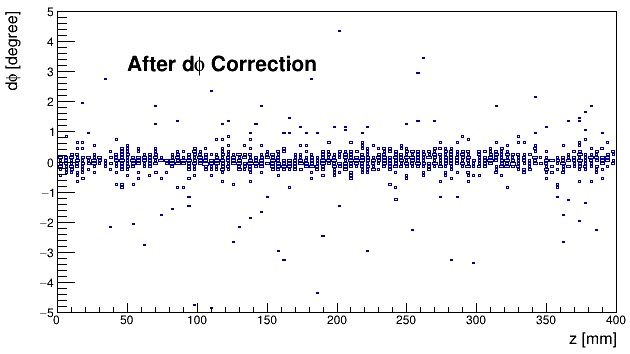}
\caption[z dependence of fiber deviation in the straight layer]{
$z$ dependence of $d\phi$ in the $\phi 3$ layer. The $z$ position is obtained from the tracking before (top) and after (bottom) the fiber position correction.  
}
\label{fiber_zure}
\end{figure}

\begin{figure}
\centering
\includegraphics[width=50mm]{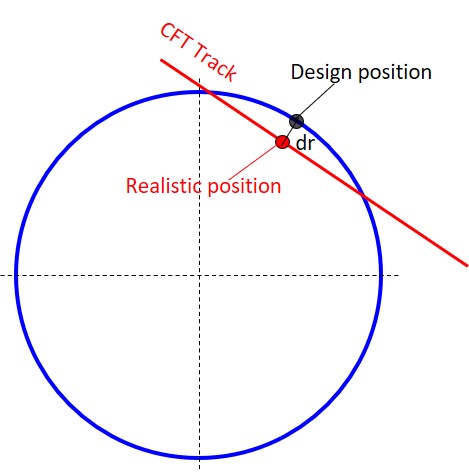}
\caption[Definition of dr]{
Definition of the deviation of the radial position ($dr$) in the $xy$ plane. $dr $ represents the difference between the desired position and the estimated position obtained through tracking in the radial direction.
}
\label{dr_zu}
\end{figure}

\begin{figure}
\centering
\includegraphics[width=70mm]{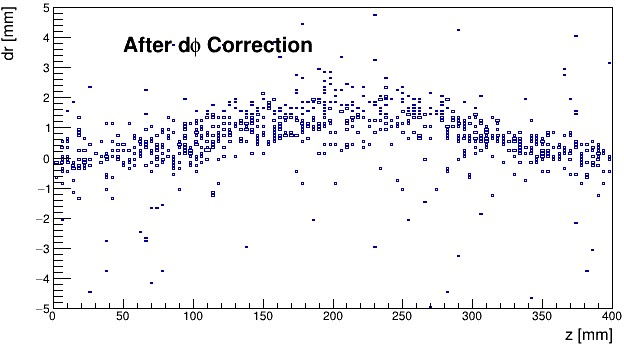}
\caption[Definition of dr]{
$dr$ distribution of a specific layer after position correction in the $d\phi$ direction. 
}
\label{dr_before}
\end{figure}

 We then describe the analysis for the fiber position correction in detail. The $\phi$ layer determines the azimuthal angle $\phi$. To represent the deviation of the fiber position, the difference in the azimuthal angle ($d\phi$) between the original designed value and the reconstructed value from the tracking is introduced, as shown in Figure \ref{dphi_zu}. If the fiber position in the analysis matches the actual fiber position, $d\phi$ should be 0. The top figure in Figure \ref{fiber_zure} shows the $z$ dependence of $d\phi$ in the $\phi 3$ layer, where the $z$ position is obtained from the tracking process. The deviation is small at both edges of the active region ($z=0$ and 400 mm), where the fibers are attached to the fiber-fixing frames, whereas the deviation increases at the center region. 
This is consistent with the characteristics of the CFT structure. The position deviation was corrected according to $d\phi$ as a function of the $z$ position. Because the $z$ dependence of the deviation is different from fiber to fiber even in the same layer, the position correction for each 10 fibers was performed. A fifth-order polynomial function was adopted to reproduce the $z$ dependence of $d\phi$ reasonably.
After the correction, the deviation was well suppressed, as shown in the bottom figure of Figure \ref{fiber_zure}. Next, the deviation in the radial direction ($dr$), as shown in Figure \ref{dr_zu},  was also verified to obtain improved fiber position.
Figure \ref{dr_before} shows the $dr$ distribution of the $\phi$ layer after correcting the position of $d\phi$. 
This deviation was also corrected to remove this correlation.
For the evaluation of the $\phi$ deviation, cosmic ray events passing near the central axis ($z$ axis) of the CFT cylinder were used because such trajectories were sensitive to the deviation of $\phi$, as shown in Figure \ref{dphi_zu}.
Conversely, for the evaluation of the $r$ deviation, tracks, apart from those in the central axis were used, as shown in Figure \ref{dr_zu}. Such a separation analysis was valid for circular-shaped detectors, such as the CFT.


The fiber position correction was performed similarly for the $uv$ layer, where the correlation between the $\phi$ and $z$ positions was calibrated using the tracking information fiber by fiber.
The deviation of the radial direction was also verified.
We iterated the tracking using the corrected fiber position and the correction of the fiber position after the tracking until the deviation became negligible.

\subsection{Angular resolution for the cosmic ray}
\begin{figure}
\centering
\includegraphics[width=75mm]{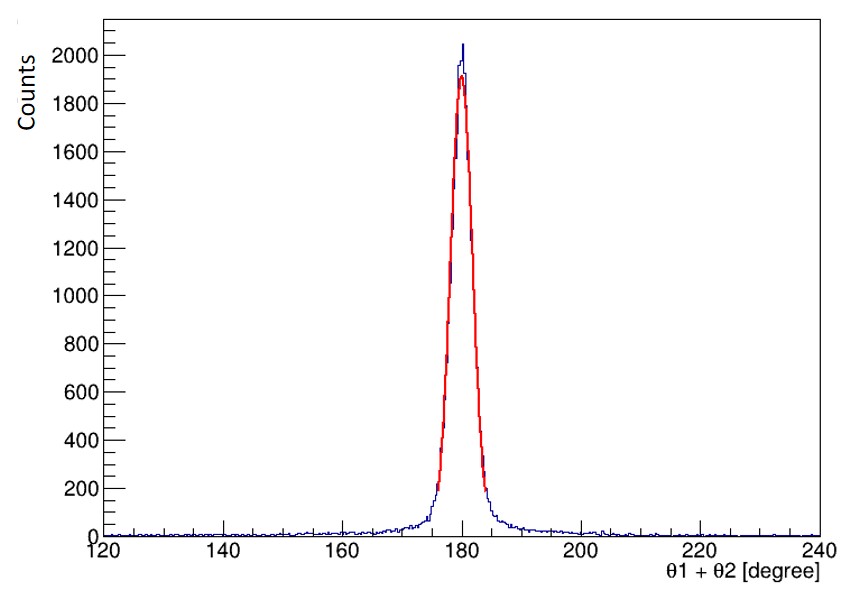}
\caption[$\theta_1 + \theta_2$ for cosmic ray]{
The $\theta_1 + \theta_2$ distribution measured using the CFT for cosmic rays.
} 
\label{thetaSum_cos}
\end{figure}

The angular resolution of the CFT for the cosmic ray was evaluated after completing the fiber position correction.
The zenith angular ($\theta$) distribution was obtained using the CFT, where 90$^{\rm o}$ in $\theta$ corresponding to the vertical direction in the laboratory frame because the $z$ axis was the central axis of the CFT. 
For the evaluation of the angular resolution of the CFT, we used the summation of both zenith angles ($\theta_{1}+\theta_{2}$), where $\theta_{1}$ and $\theta_{2}$ were obtained through the top eight layers and the bottom eight layers of the CFT, respectively.
This $\theta_{1}+\theta_{2}$ should be 180$^{\rm o}$ for the same straight track, as shown in Figure \ref{thetaSum_cos}. 
The spread of $\theta_{1}+\theta_{2}$ ($\sigma_{\theta_{1}+\theta_{2}}$) was obtained as 1.84$^{\rm o}$ in $\sigma$.
The angular resolution ($\sigma_{\theta}$) obtained through the eight-layer tracking was 1.30$\pm$ 0.01$^{\rm o}$ assuming that the spread in $\theta_{1}+\theta_{2}$ was expressed as follows: $\sigma_{\theta_{1}+\theta_{2}} = \sqrt{2}\sigma_{\theta}$.
This result satisfied the requirement of 2$^{\rm o}$ in $\sigma$ for the $\Sigma p$ scattering experiment.
The mean value of the $\theta_{1} + \theta_{2}$ peak was 179.99$\pm$ 0.01$^{\rm o}$, as expected.


\section{Performance evaluation of CATCH using a proton beam}
\begin{figure}
\centering
\includegraphics[width=60mm]{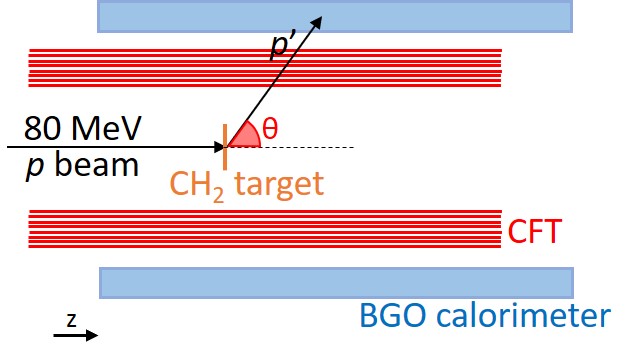}
\caption[Setup for the $pp$ and $p$C scattering experiments]{
Setup for the $pp$ and $p$C scattering experiments with a CH$_2$ target and CATCH.}
\label{ch2tgt}
\end{figure}

To evaluate the performance of CATCH, proton--proton ($pp$) and proton--carbon ($p$C) scattering experiments were performed in January 2017 at the Cyclotron Radio Isotope Center (CYRIC) in Tohoku University.
An 80 MeV proton beam was irradiated on a polyethylene (CH$_{2}$) target of 800 $\mu$m thickness installed inside CATCH.
The experimental setup is shown in Figure \ref{ch2tgt}.
The scattered protons were detected using CATCH.
The target position was changed to three different positions ($z$=54, 126, and 208 mm) to determine the $z$ dependence of the performance along the sensitive area of 400 mm in the $z$ direction.
The trigger for the data acquisition was the coincidence between the inner two $\phi$ layers and the RF signal of 50 MHz synchronized with the beam incident timing from the accelerator.
The experimental conditions are listed in Table \ref{condition}.

\begin{table}
\begin{center}
\small
  \caption{Experimental condition of the CATCH test with a CH$_2$ target in CYRIC.}

  \begin{tabular}{|c|c|} \hline
   setup and conditions & Polyethylene (CH$_2$) target  \\ \hline 
     Target Position  & 
  \begin{tabular}{lc|}
$z$ = $54, 126, 208$ mm \\ 
change position with one target
  \end{tabular} \\ \hline
   Target thickness   & $800 \mu$m  \\ \hline
  Beam particle   & proton  \\ \hline
   Beam energy   & $78.5$ MeV  \\ \hline
   Beam size   & About $5(x)\times10(y)$  mm$^2$  \\ \hline
   Beam intensity   & About $0.05$ nA  \\ \hline
   Trigger   &  CFT ($\phi$1 $\times$ $\phi$2) $\times$ RF \\ \hline
   Beam time   & $3.5$ days  \\ \hline
  \end{tabular}
    \label{condition}
\end{center}
\end{table}

\subsection{Overview of the performance evaluation of CATCH}
We first describe the procedure for evaluating the performance of CATCH as a system for measuring differential cross sections.

To identify the $pp$ and $p$C elastic scattering events, we require  information regarding both the scattering angle and the kinetic energy of the scattered proton.
The scattering angle of the proton is obtained using the CFT, as explained in Section 4.
The energy calibrations of the CFT and the BGO calorimeter are performed by comparing the energy of the scattered proton calculated from the scattering angle $\theta$ and the measured energy deposits on each detector for the $pp$ and $p$C scattering events.
For this purpose, the CFT is necessary; it also affects the accuracy of the calibration directly.
Therefore, the tracking performance of the CFT is evaluated first.
After the energy calibration, the energy resolutions of the BGO calorimeter and the CFT are evaluated.

To identify the $pp$ elastic scattering, we require that the relationship between the kinetic energy and the scattering angle is consistent with the scattering kinematics.
Here, we define an index $\Delta E_{kin}$, which represents the consistency in the $pp$ elastic scattering defined as follows:
\begin{equation}
\Delta E_{kin} = E_{measure} - E_{calc} (\theta),
\end{equation}
where $E_{measure}$ represents the energy measured using the BGO calorimeter and the CFT, and $E_{calc} (\theta)$ represents the calculated energy for the scattering angle $\theta$ based on the $pp$ elastic scattering kinematics.
This index can be applied to other scattering reactions, such as the $p$C and $\Sigma p$ scatterings using each kinematical relation between the energy and the scattering angle.
To identify the $\Sigma p$ scattering event from the background reaction, the resolution of $\Delta E_{kin}$ is required to be higher than 7 MeV in $\sigma$.
The $\Delta E_{kin}$ resolution is evaluated through the $pp$ scattering events.
Finally, the differential cross section is derived as the total performance of CATCH.

\subsection{Angular resolution in the CFT}
\begin{figure}
\centering
\includegraphics[width=70mm]{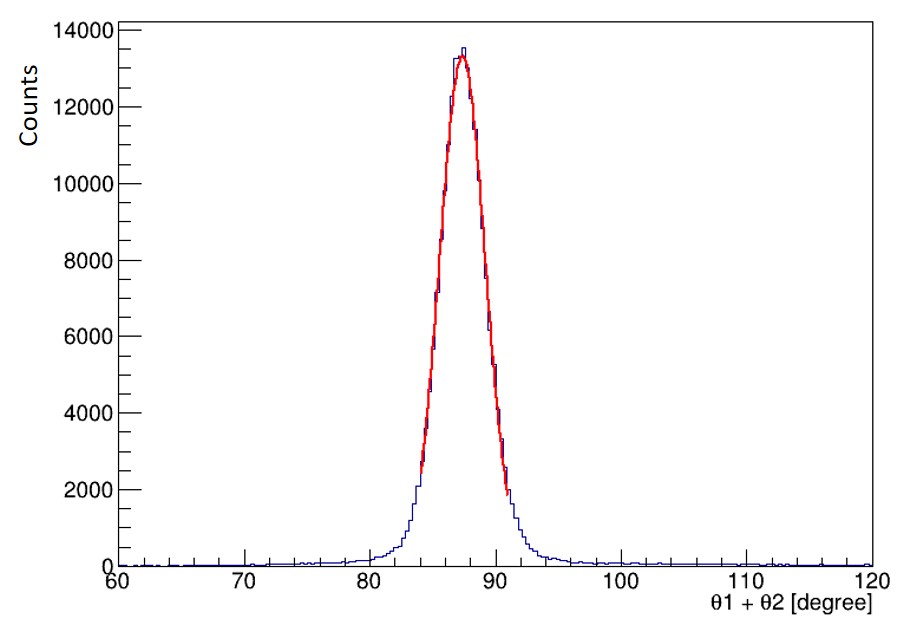}
\caption[Schema of the summation of 2 $\theta$ from the $pp$ scattering experiment]{The opening angle for the $pp$ scattering events. It was obtained by adding the two scattering angles $\theta_{1}$ and $\theta_{2}$ for two protons.
}
\label{theta_wa_zu}
\end{figure}

\begin{figure}
\centering
\includegraphics[width=70mm]{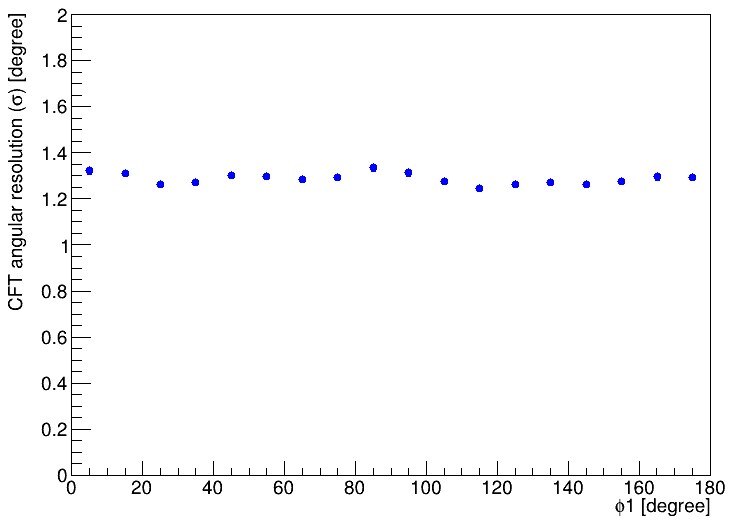}
\caption[Summation of the measured 2 $\theta$ from the $pp$ scattering experiment]{The $\phi$ dependence of the CFT angular resolution ($\sigma_\theta$).  
The angular resolution was obtained uniformly over the $\phi$ direction.
}
\label{theta_wa}
\end{figure}

In the analysis, the fiber position correction explained in Section 4 has already been applied.
In the $pp$ scattering event, two protons are emitted from the target. 
\color{black}
The opening angle between the two protons is slightly smaller than 90$^{\rm o}$ owing to the relativistic effect. 
\color{black}
Figure \ref{theta_wa_zu} shows the opening angle distribution obtained by adding both scattering angles, $\theta_{1}$ and $\theta_{2}$, for two protons.
Assuming that the angular resolution is similar for both tracks, i.e., the same assumption for the cosmic ray measurement, the spread of the opening angle ($\theta_{1}+\theta_{2}$) should be $\sqrt{2}\sigma_{\theta}$, where $\sigma_{\theta}$ represents the angular resolution of the single track.
The angular resolution $\sigma_{\theta}$ was 1.27$\pm$ 0.01$^{\circ}$.
It is also crucial to verify the uniformity of the angular resolution for all the azimuthal ($\phi$) regions.
This uniformity was confirmed by determining the $\phi$ dependence of $\sigma_{\theta}$, as shown in Figure \ref{theta_wa}.
From a simulation in which the misalignment of fiber positions is not considered, the angular resolution was estimated as $\sigma_{\theta}=1.04$$^{\circ}$ as a result of similar evaluations.
Although there is an insignificant difference between the angular resolution data and the simulation data, the obtained angular resolution satisfies the requirement of 2$^{\circ}$.

\subsection{Time resolution of the CFT}

\begin{figure}
\centering
\includegraphics[width=70mm]{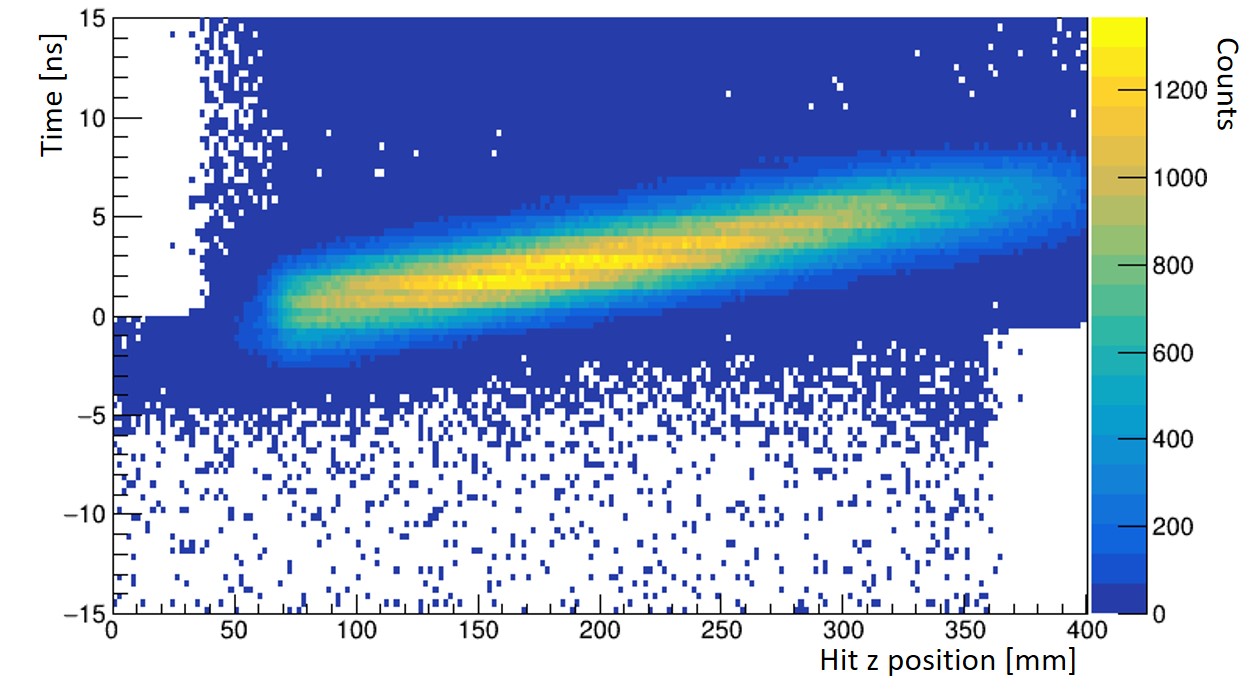}
\caption[z position dependence of the CFT time distribution]{
$z$ position dependence of the CFT time distribution of the $v$4 layer. 
The correlation corresponds to the propagation time of the light in the fiber.
}
\label{ctime_track_z}
\end{figure}

The time resolution of the CFT should be evaluated as one of the most crucial performance evaluations of the CFT. This is because the time resolution determines the minimum time gate for the truly triggered event under the high counting rate in the $\Sigma p$ scattering experiment.
As a result of the analysis, the transmission time of the scintillation photon in the scintillation fiber affects the timing resolution in the CFT.
Figure \ref{ctime_track_z} shows the correlation between the time response of the CFT and the $z$ position of the charged particle obtained from the CFT.
There is a clear correlation between these values corresponding to the propagation time of the light in the fiber.
This $z$ dependence in the time measurement was corrected using a linear function.
The results of the time resolutions without and with the $z$ position correction are listed in Table \ref{time_sum}.
The time resolution higher than 1.8 ns in $\sigma$ was achieved in all the layers.

\begin{table}
\begin{center}
\small
  \caption{Summary of the CFT time resolution ($\sigma$) for each layer in ns.}
\scalebox{0.8}[0.8]{
  \begin{tabular}{|c|c|c|c|c|c|c|c|c|} \hline
    $\sigma_{time}$ ns   & $u$1 & $v$2 & $u3$ & $v$4    & $\phi$1 & $\phi$2 & $\phi$3 & $\phi$4  \\ \hline 
  not corrected & 2.02 & 2.24 & 2.39 & 2.47      & 1.98 & 1.98 & 1.86 & 1.83  \\ \hline
  z corrected   & 1.57 & 1.68 & 1.63 & 1.68      & 1.76 & 1.72 & 1.54 & 1.56 \\ \hline
  \end{tabular}
}
    \label{time_sum}
\end{center}
\end{table}

\subsection{Energy calibration of the BGO calorimeter}
In this section, we describe the energy calibration of the BGO calorimeter using the kinematic relation for the $pp$ and $p$C scattering experiments.
As explained in Section 3.2, the waveform data of the shaped signal with an integral circuit were recorded using a flash ADC with a sampling frequency of 33 MHz.
The pulse height and the timing of each waveform were obtained through the template fitting method.

\begin{figure}
\centering
\includegraphics[width=70mm]{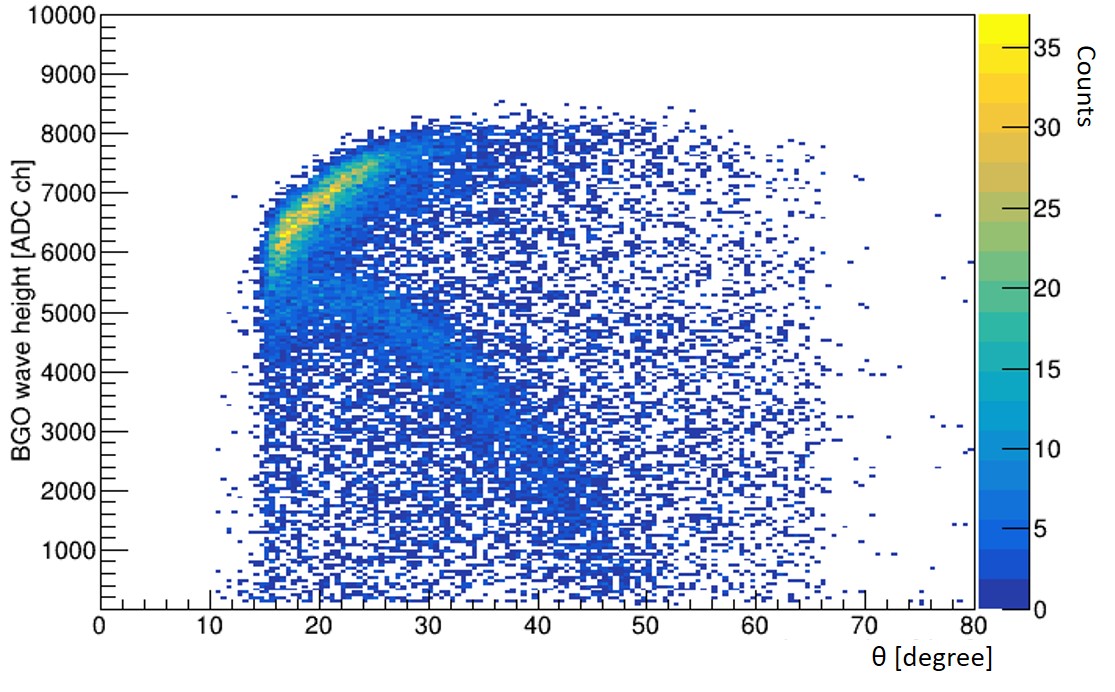}
\caption[Correlation between the BGO wave height and $\theta$]{
Correlation between the scattering angle $\theta$ and the BGO pulse height for protons. 
}
\label{bgo_ht}
\end{figure}

\begin{figure}
\centering
\includegraphics[width=65mm]{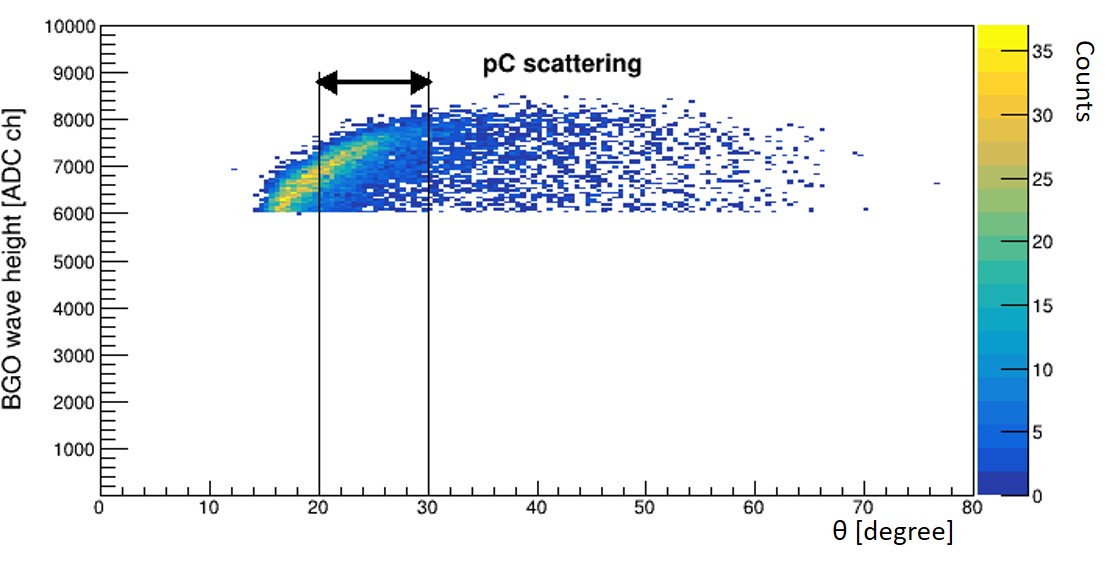}
\includegraphics[width=65mm]{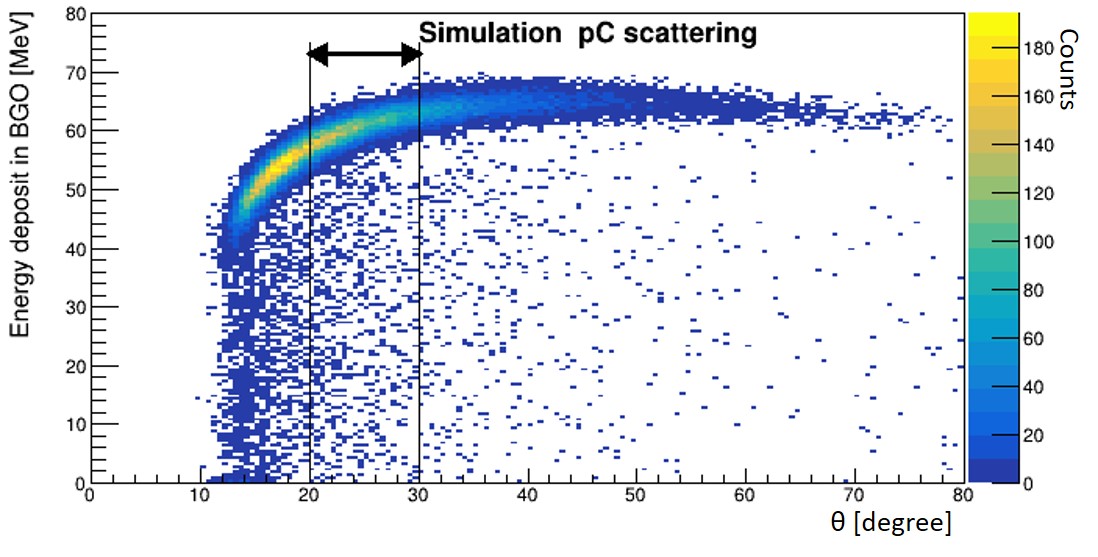}
\caption[Simulated energy deposit in BGO calorimeter from $pp$ and $p$C scattering]{
Angular dependences of the pulse height in the data (top) and the simulated energy deposit (bottom) in the BGO calorimeter for the $p$C scattering events. By selecting a large pulse height region of the BGO calorimeter, $p$C scattering events were selected. 
}
\label{bgo_pC}
\end{figure}

\begin{figure}
\centering
\includegraphics[width=65mm]{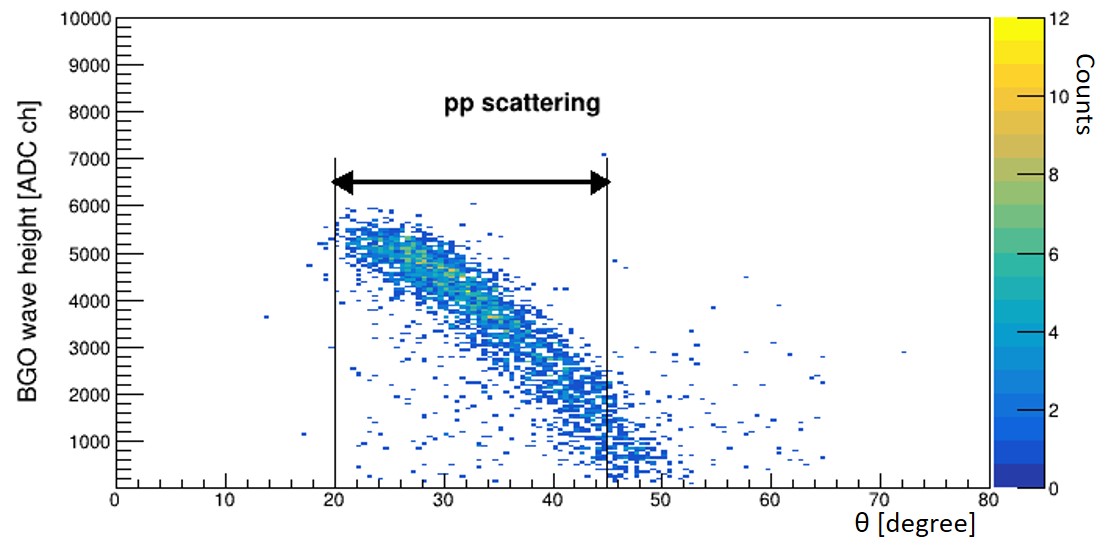}
\includegraphics[width=65mm]{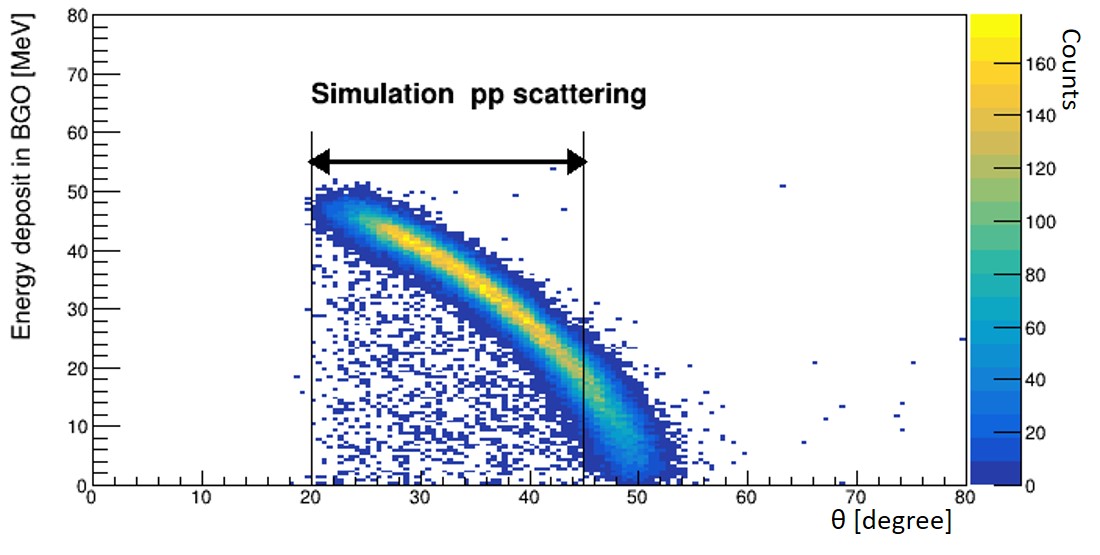}
\caption[Selected BGO wave height for $pp$ and $p$C scattering]{
Angular dependences of the pulse height in the data (top) and the simulated energy deposit (bottom) in the BGO calorimeter for the $pp$ scattering events.  
In the data analysis shown in the top figure, the back-to-back hit correlation in the $\phi$1 layer was required to select the $pp$ scattering event.
}
\label{bgo_pp}
\end{figure}

The energy of protons incident on the BGO calorimeter has a one-to-one correlation with the scattering angle $\theta$ because the $pp$ and $p$C elastic scatterings are two-body reactions.
However, protons pass through the CFT before they arrive at the BGO calorimeter.
Therefore it is necessary to consider the energy loss in the CFT to obtain the incident proton energy at the BGO surface. 
Figure \ref{bgo_ht} shows the correlation between the scattering angle $\theta$ and the BGO pulse height for protons. 
The two loci corresponding to the $pp$ and $p$C elastic scatterings are confirmed in this figure.
These angular dependences of the pulse height were compared with the simulated relation between the angle $\theta$ and the energy deposit in the BGO calorimeter, as shown in Figure \ref{bgo_pC} and \ref{bgo_pp} for the $p$C and $pp$ scattering events, respectively. 

\begin{figure}
\centering
\includegraphics[width=70mm]{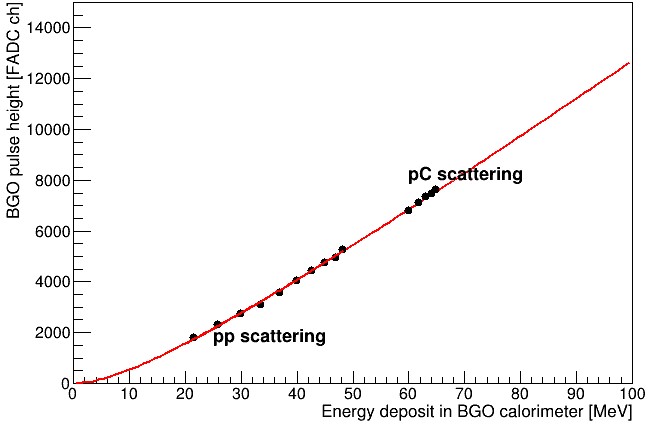}
\caption[Correlation between the measured BGO wave height and the simulated energy deposit]{
Correlation between the measured BGO pulse height and the simulated energy deposit for the $pp$ and $p$C scatterings. 
The red line shows the energy calibration function of Equation (\ref{eq_calib}).
}
\label{bgo_calib}
\end{figure}

\color{black}
The $p$C elastic scattering events were selected in a large pulse height region of the BGO calorimeter, as shown in the top figure of Figure \ref{bgo_pC}.
\color{black}
The $pp$ scattering events can be selected more accurately by requiring the back-to-back hit correlation in the $\phi$1 layer corresponding to the two protons in the $pp$ scattering event.
\color{black}After this selection\color{black}, the top figure of Figure \ref{bgo_pp} shows the correlation between the scattering angle and the BGO pulse height for the $pp$ scattering event.
To avoid the misidentification resulting from the overlap between the $pp$ and $p$C scattering events, the angular regions of 20 $\le \theta$ (degree) $\le$ 45 and the 20 $\le \theta$ (degree) $\le$ 30 were selected for the $pp$ and $p$C scattering events, respectively.
By comparing the measured pulse height of the BGO calorimeter to the simulated energy (bottom figures of Figures \ref{bgo_pC} and \ref{bgo_pp}) for the same scattering angle $\theta$, the relation between the BGO pulse height and the energy deposit was obtained as shown in Figure \ref{bgo_calib}. 
\color{black}
Light yield saturation with a large energy deposit per unit length in the inorganic scintillator could account for the correction.
A phenomenological model describes this correlation by the following equation \cite{Avdeichikov:1993}:
\color{black}
\begin{equation}\label{eq_calib}
PH = a \times E - b\times \ln 
\left(
\frac{E+b}{b}
\right)
,
\end{equation}
 where $PH$ and $E$ correspond to the pulse height and the energy deposit in the BGO calorimeter, respectively.
 The parameters $a$ and $b$ were obtained by fitting the data points in Figure \ref{bgo_calib} with this function.
 The BGO energy calibration was performed for all BGO segments using the method presented above.
 
 \subsubsection{The BGO energy resolution}
\begin{figure}
\centering
\includegraphics[width=70mm]{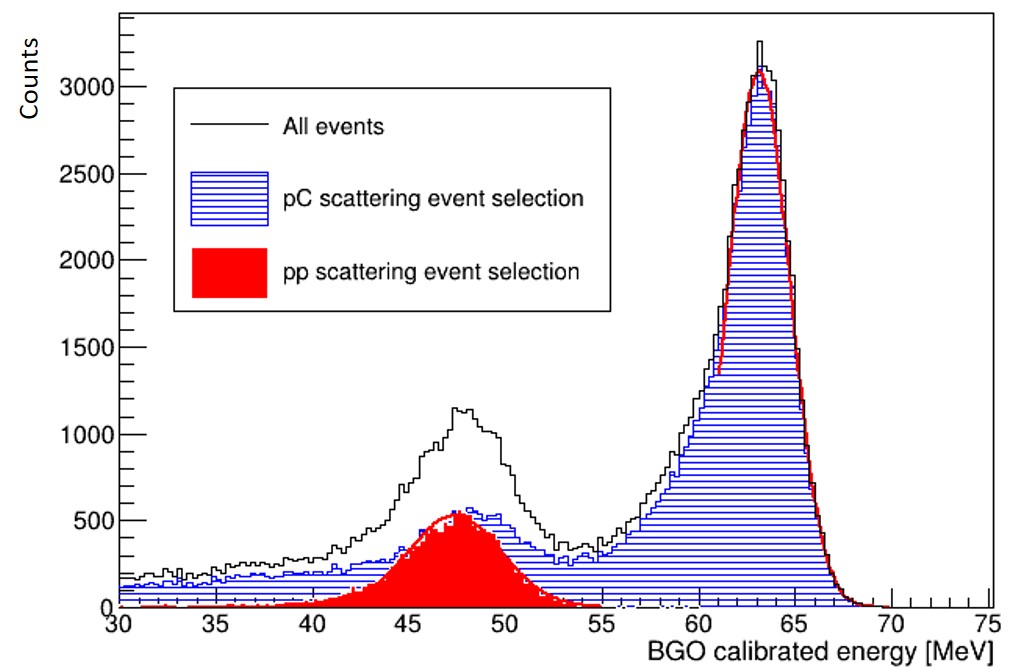}
\caption[BGO energy selected in scattering angular range of 25 $\pm$ 2.5 degrees]{
Energy distribution measured using the BGO calorimeter for the scattering angle of 25 $\le$ $\theta$ (degree) $\le$ 30. 
}
\label{EBGO}
\end{figure}

\begin{figure}
\centering
\includegraphics[width=70mm]{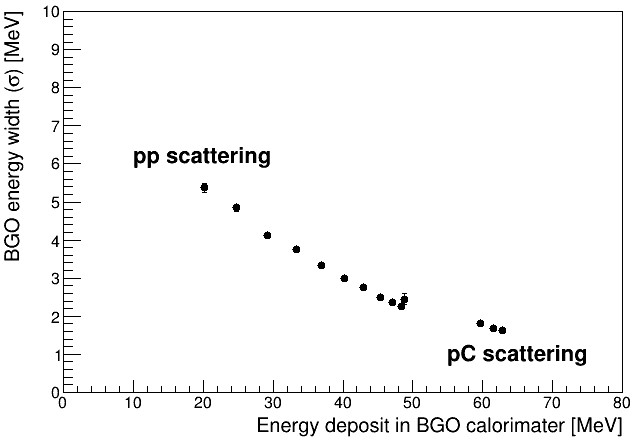}
\caption[$\theta$ dependence of the BGO energy resolution]{
Energy dependence of the BGO energy width estimated from the $pp$ and $p$C scattering events with different angular regions.
}
\label{EBGOline}
\end{figure}

 In this section, the energy resolution of the BGO calorimeter is discussed.
 It is difficult to evaluate the intrinsic energy resolution of the BGO calorimeter from this experiment. This is because the energy spread in the BGO calorimeter is affected by the energy struggling in the CFT.
 Here, we approximately evaluate the energy resolution by comparing the data and the simulation considering the energy resolution of the BGO calorimeter, and we determine whether the assumed resolution is obtained. Figure \ref{EBGO} shows the BGO energy distribution for the scattering angle of 25 $\le$ $\theta$ (degree) $\le$ 30. The peaks at 63 MeV and 48 MeV correspond to the $p$C and $pp$ scatterings, respectively. Because there are contributions of the $p$C inelastic scattering, there exists a tail toward the low-energy side of the $p$C elastic scattering peak.
 Therefore, the width of the $p$C elastic scattering peak was estimated by fitting the energy spectrum of the asymmetric region to exclude the contribution of the inelastic scattering.
 The energy dependence of the BGO energy width ($\sigma$) was estimated from the $pp$ and $p$C scattering events with different angular regions, as shown in Figure \ref{EBGOline}. In this energy spread, the contributions from the energy resolution of the BGO and the energy struggling in the CFT are included. This width was compared to the simulation involving the same experimental setup, including the CFT, by replacing the intrinsic energy resolution of the BGO with an assumption of energy dependence of $\frac{\sigma_{E_{BGO}}}{E} \propto \frac{1}{\sqrt{E}}$ for the BGO energy resolution. The experimental width could be approximately reproduced through the simulation, where the BGO energy resolution was 1.5\% for 80 MeV protons. Therefore, the energy resolution with such levels might be obtained, and it is concluded that the energy resolution is better than the required value of 3\% for 80 MeV protons.

 \subsection{Energy calibration of the CFT}
 The energy calibration of the CFT was performed by comparing a measured ADC value to a simulated energy deposit for each scattering angle $\theta$, as mentioned in the calibration of the BGO calorimeter. The energy deposits for protons with similar scattering angles are different in the $\phi$ and $uv$ layers owing to the layer structure. They are explained separately in the following section.

  \subsubsection{Normalization of ADC data of the CFT}
 To ensure that the operating condition was uniform among all the MPPCs, the amplification factor of each MPPC was set to be as uniform as possible by adjusting the operation voltage one after the other through the VME-EASIROC board. In the ideal case, the energy calibrations of the CFT should be performed sequentially for each fiber to manage the possible gain difference.
However, this approach is not realistic owing to the huge numbers of segments and insufficient statistics of the calibration data for each fiber.
Therefore, we normalized the ADC values for all fibers in each layer, and the calibration was performed for the normalized ADC for each layer.
 The mean value of the ADC distribution for the $pp$ elastic scattering of 20 $\le$ $\theta$ (degree) $\le$ 30 was normalized to be five for all fibers. 
 This normalized ADC was calibrated to the energy of each layer. 

 \subsubsection{Energy calibration for the $\phi$ layers}
 
\begin{figure}
\centering
\includegraphics[width=55mm]{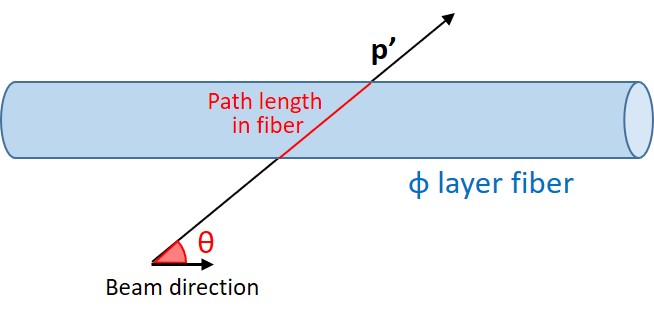}
\caption[Correlation between $\theta$ and the energy deposit in the CFT during simulation]{ Schematic illustration of the relationship between a path length in a fiber and a scattering angle $\theta$ for the $\phi$ layer. 
}
\label{path_phi}
\end{figure}

\begin{figure}
\centering
\includegraphics[width=65mm]{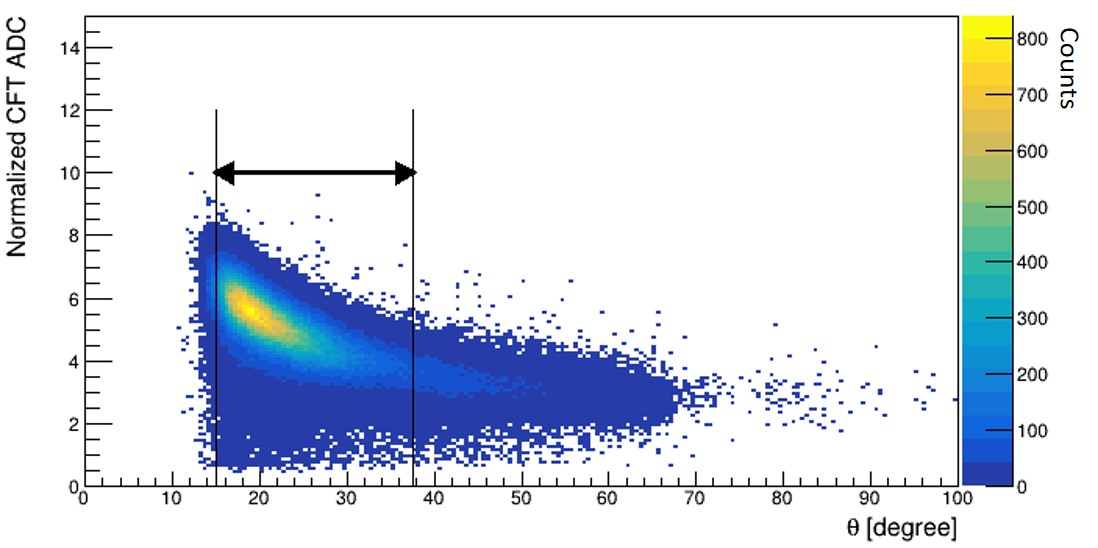}
\includegraphics[width=65mm]{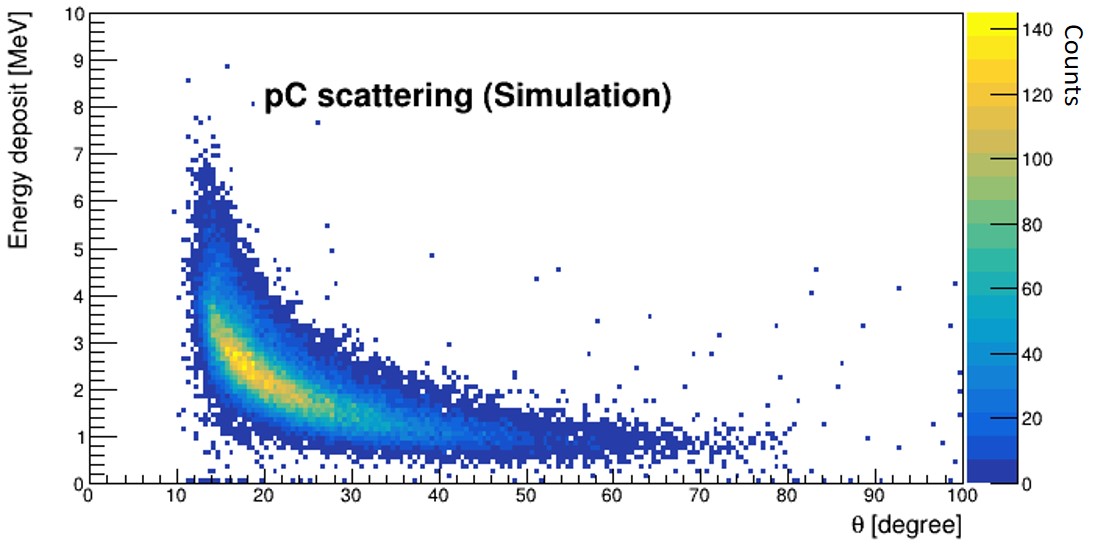}
\caption[Correlation between the number of detected photons and the simulated energy deposit in the CFT]{
(Top) Scatter plot between $\theta$ and the normalized ADC for the $\phi$ layer.
 (Bottom) Simulated angular dependence of the energy deposit in the CFT $\phi$ layer for $p$C scattering.
}
\label{dECFT_phisim}
\end{figure}

\begin{figure}
\centering
\includegraphics[width=70mm]{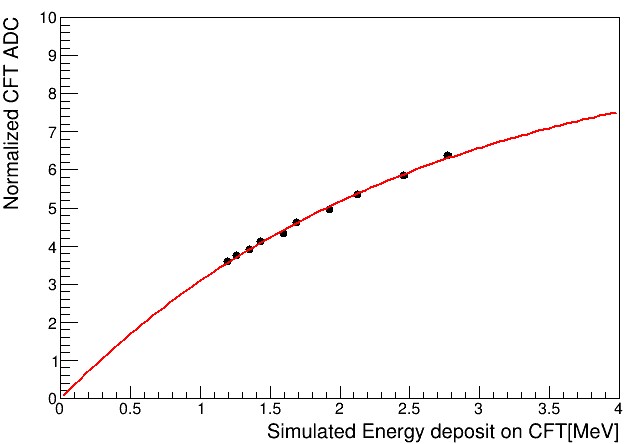}
\caption[Correlation between the number of detected photons and the simulated energy deposits in the CFT]{
Correlation between the normalized ADC and the simulated energy deposit for the $\phi$ layer in the CFT.
The red line shows the calibration function obtained by fitting the data points to the function defined in Equation (\ref{eq_calib_mppc}).
}
\label{dECFT_comp}
\end{figure}

 Because the $\phi$ layer's fiber is placed parallel to the direction of the beam, the path length inside the fiber depends on the scattering angle $\theta$ of the proton, as shown in Figure \ref{path_phi}. Specifically, for the proton from the $p$C scattering experiment, the energy deposit in the fiber mainly depends on the path length because the energy of the proton does not significantly depend on the scattering angle. The top and bottom figures of Figure \ref{dECFT_phisim} show the correlations  between the scattering angle and the normalized ADC (energy deposit for the simulation) of the $p$C scattering for data and the simulation, respectively. The normalized ADC and the energy deposit vary depending on the scattering angle. Similar to the energy calibration of the BGO calorimeter, the normalized ADC of the CFT was compared to the simulated energy deposit for each scattering angle, as shown in Figure \ref{dECFT_comp}. As shown in the figure, it is clear that the normalized ADC becomes saturated for the large energy deposit owing to the saturation effect of the MPPC.
 The output of the MPPC is proportional to the number of activated APD pixels in the MPPC by incident photons.
 As the number of incident photons increases, multiple photons are incident on the same APD  pixel, thereby resulting in the non-linearity of the output signals.
The relationship between the number of the activated APD pixels and the energy deposit, which is assumed to be proportional to the number of the incident photons for simplicity, is expressed as follows:
\begin{equation}\label{eq_calib_mppc}
N_{detected} = N_{pixel} \left\{
1 - \exp \left(
\frac{-b\times dE}{N_{pixel}}
\right)
\right\}
,
\end{equation}
 where $N_{detected}$ and $N_{pixel}$ represent the numbers of activated APD pixels and the effective pixels covering the fiber edge, respectively, and  $b \times dE$ corresponds to the incident photon number expressed from a parameter $b$, i.e., a coefficient from the energy deposit ($dE$) to the photon number.
By fitting the correlation presented in Figure \ref{dECFT_comp} to this function, the energy calibration function is obtained.
As a result, the effective pixel number $N_{pixel}$ was 200, which was almost consistent with the pixel number of 180 covered geometrically by a fiber of 0.75 mm in diameter.
Similar calibration functions were obtained for all $\phi$ layers, and this was reasonable because the operational condition was almost common for all MPPCs.

 \subsubsection{Energy calibration for the $uv$ layers}
\begin{figure}
\centering
\includegraphics[width=70mm]{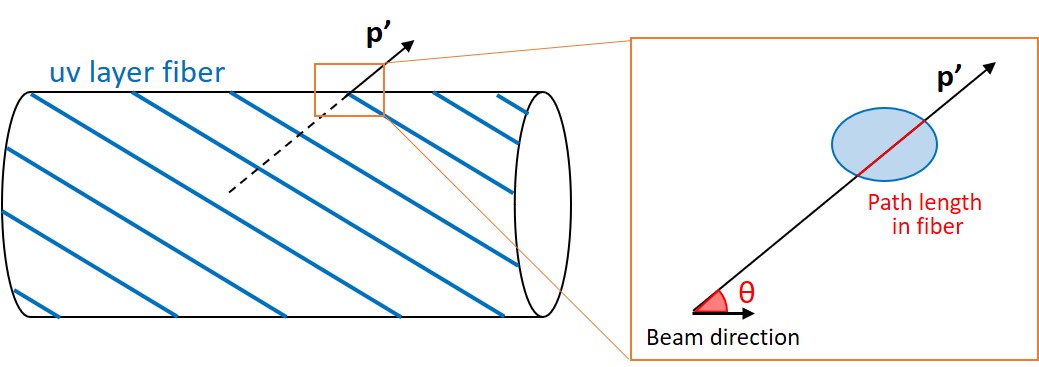}
\caption[Cross sectional shape of the fiber section on the plane of the scattered proton.]{
Cross sectional shape of the CFT $uv$ fiber for the scattering plane of the scattered proton.
The cross section of the $uv$ layer becomes elliptical, and the angular dependence of the path length in the $uv$ fiber is much smaller compared to that of the $\phi$ fiber.
}
\label{pathUV}
\end{figure}

Contrary to the $\phi$ layers, the energy deposit in the $uv$ layers does not depend on the scattering angle because the cross section of the $uv$ layer becomes elliptical, as shown in Figure \ref{pathUV}.
Therefore, it was difficult to make the correlation between the simulated energy deposit and the normalized ADC, as shown in Figure \ref{dECFT_comp} for the $\phi$ layers.
From the study of the $\phi$ layers, the effective pixel number is almost similar because it is determined from the geometrical point of view between the MPPC and the fiber size.
Therefore, we also used $N_{eff}$=200 (the typical value for the $\phi$ layers) for the $uv$ layers.
The parameter $b$ was then obtained, so that the energy deposit for $\theta$=25$^{\rm o}$ corresponded to the estimated value from the simulation.
Here, the ambiguity of the energy deposit in one fiber depends strongly on the energy of the incident proton.
To minimize the uncertainty of the incident proton's energy, the higher energy proton emitted to the forward angular region of  $\theta$=25$^{\rm o}$ was used in this calibration.

 \subsubsection{Energy resolution of the CFT}
\begin{figure}
\centering
\includegraphics[width=70mm]{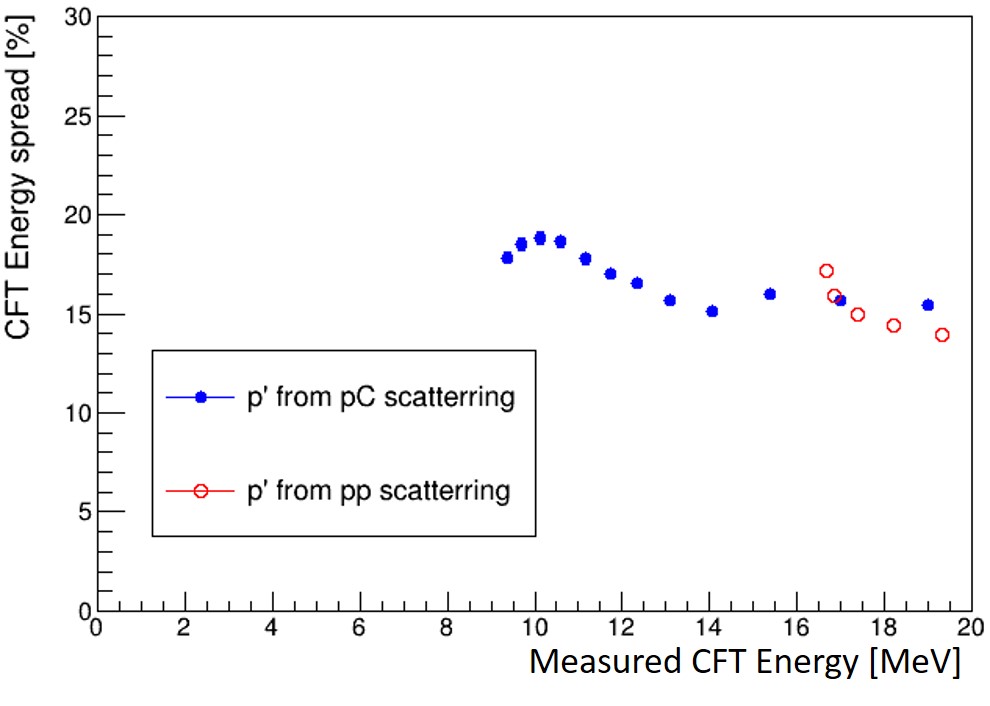}
\caption[Measured energy spread of the CFT ]{
Energy dependence of the energy resolution ($\sigma / E$) of the CFT obtained from the $pp$ and $p$C scattering experiments.}
\label{cft_reso}
\end{figure}

 The energy deposit in the CFT can be measured by summing up the energy deposits in all the CFT layers.
The energy resolution for the total energy loss in the CFT was estimated from the width in the total $dE$ spectrum for the $pp$ and $p$C scattering experiments.
The energy dependence of the resolution can be roughly estimated by changing the scattering angle of the proton, as shown in Figure \ref{cft_reso}. 
There were only small energy dependences in the energy resolution, and typically 20\% was obtained for 10 MeV energy deposits in the CFT.
This energy resolution is sufficient to separate the proton from $\pi$ using the $\Delta E$-$E$ method from the partial energy deposit in the CFT and the total energy deposit in the BGO calorimeter. This is because the energy deposit for $\pi$ is estimated to be approximately 1.5 MeV \cite{Miwa:2021}.

 \subsection{Kinematic identification of scattering events}
\begin{figure}
\centering
\includegraphics[width=70mm]{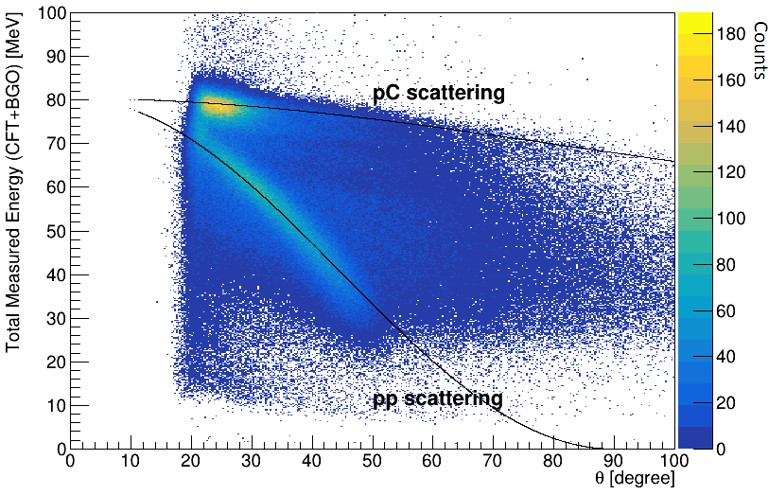}
\caption[Scatter plot of the scattering angle $\theta$ and the energy of p']{
Scatter plot between the scattering angle $\theta$ measured using the CFT and the kinetic energy of the proton measured using the CFT and the BGO calorimeter. 
The solid lines show the kinematic relationship between the $pp$ and $p$C elastic scatterings.
}
\label{scat_plot_ch2}
\end{figure}

\begin{figure}
\centering
\includegraphics[width=65mm]{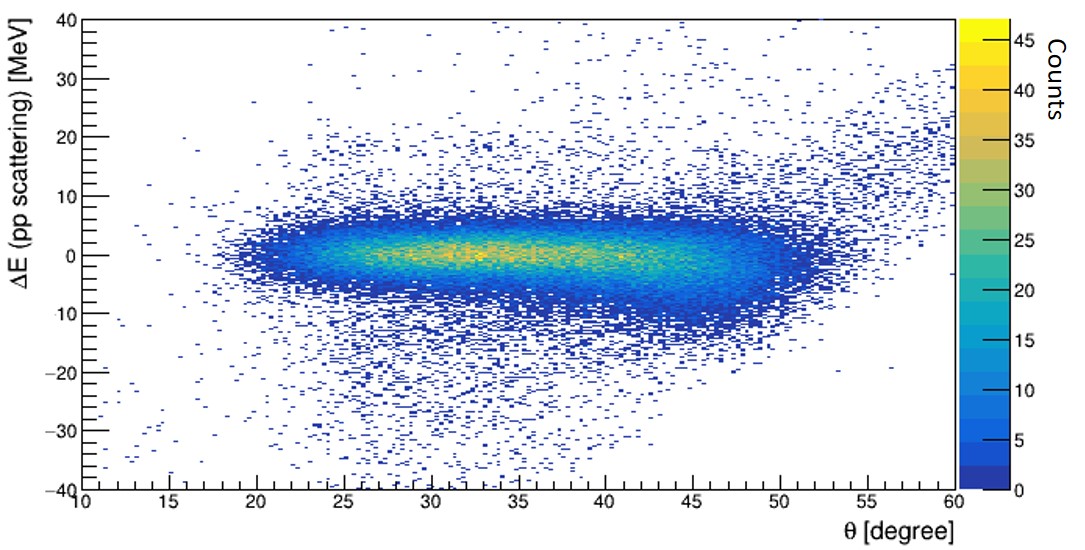}
\includegraphics[width=65mm]{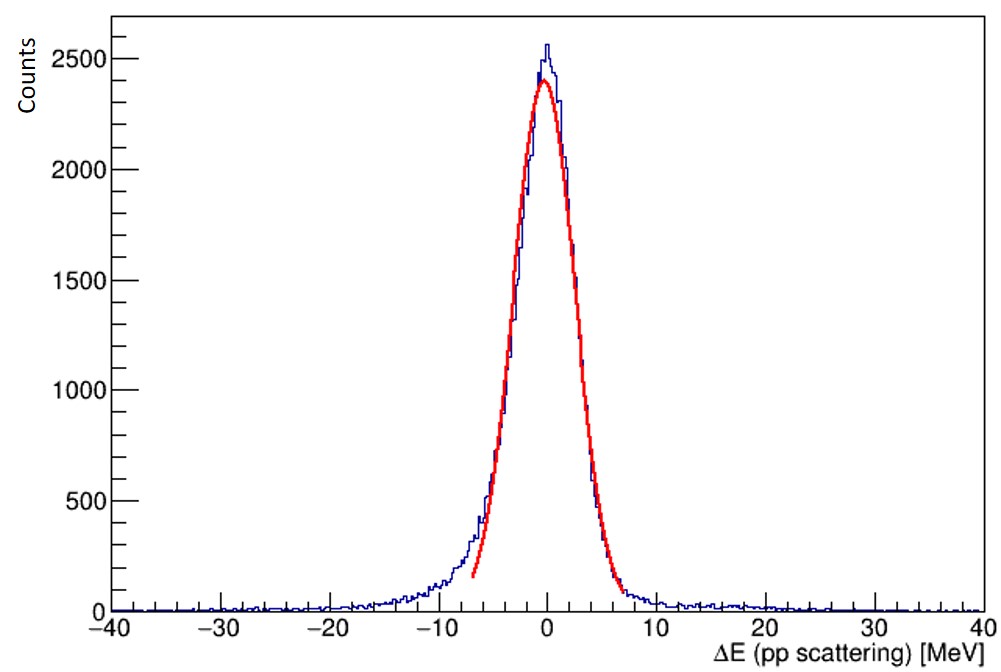}
\caption[Correlation between the $\Delta E$ and $\theta$ selected]{
(Top) Correlations between the $ \Delta E_{kin} $ and the scattering angle $\theta$ for the $pp$ scatterings. (Bottom) $\Delta E_{kin}$ distribution for the $pp$ scatterings for the entire $\theta$ region.}
\label{deltaEselect}
\end{figure}

 Because the energy calibrations of the CFT and the BGO calorimeter were performed, the kinetic energy of the scattered proton was reconstructed by adding the energy deposits in two detectors. Figure \ref{scat_plot_ch2} shows the scatter plot between the scattering angle $\theta$ measured using the CFT and the kinetic energy of the proton measured using the CFT and the BGO calorimeter. The loci corresponding to the $pp$ and $p$C scattering experiments are identified. The kinematic lines of the $pp$ and $p$C elastic scatterings are also overlaid in Figure \ref{scat_plot_ch2}.
Although the CFT and BGO were calibrated separately, the reconstructed correlations are well consistent with these kinematic lines. To identify the elastic scattering, we introduce the index $\Delta E_{kin}$ defined in Equation (1), which represents the consistency between the measured energy and the calculated energy from the scattering angle for each scattering kinematic.
This $\Delta E_{kin}$ corresponds to the difference between the measured energy and the kinematic lines presented in Figure \ref{scat_plot_ch2}. The top figure of Figure \ref{deltaEselect} shows the scatter plots between the $\Delta E_{kin}$ and $\theta$ values for the $pp$ elastic scattering.
The bottom figure of Figure \ref{deltaEselect} shows the $\Delta E_{kin}$ distribution for the $pp$ elastic scattering, and the width ($\sigma$) was 2.8 MeV for the entire $\theta$ region. This performance satisfies the requirement of 7 MeV for the identification of $\Sigma p$ scattering \cite{Miwa:2021}.
The $pp$ scattering event was identified by selecting the $\Delta E_{kin}$ peak in the $\pm 3\sigma$ region.

 \subsection{CFT tracking efficiency}
\begin{figure}
\centering
\includegraphics[width=70mm]{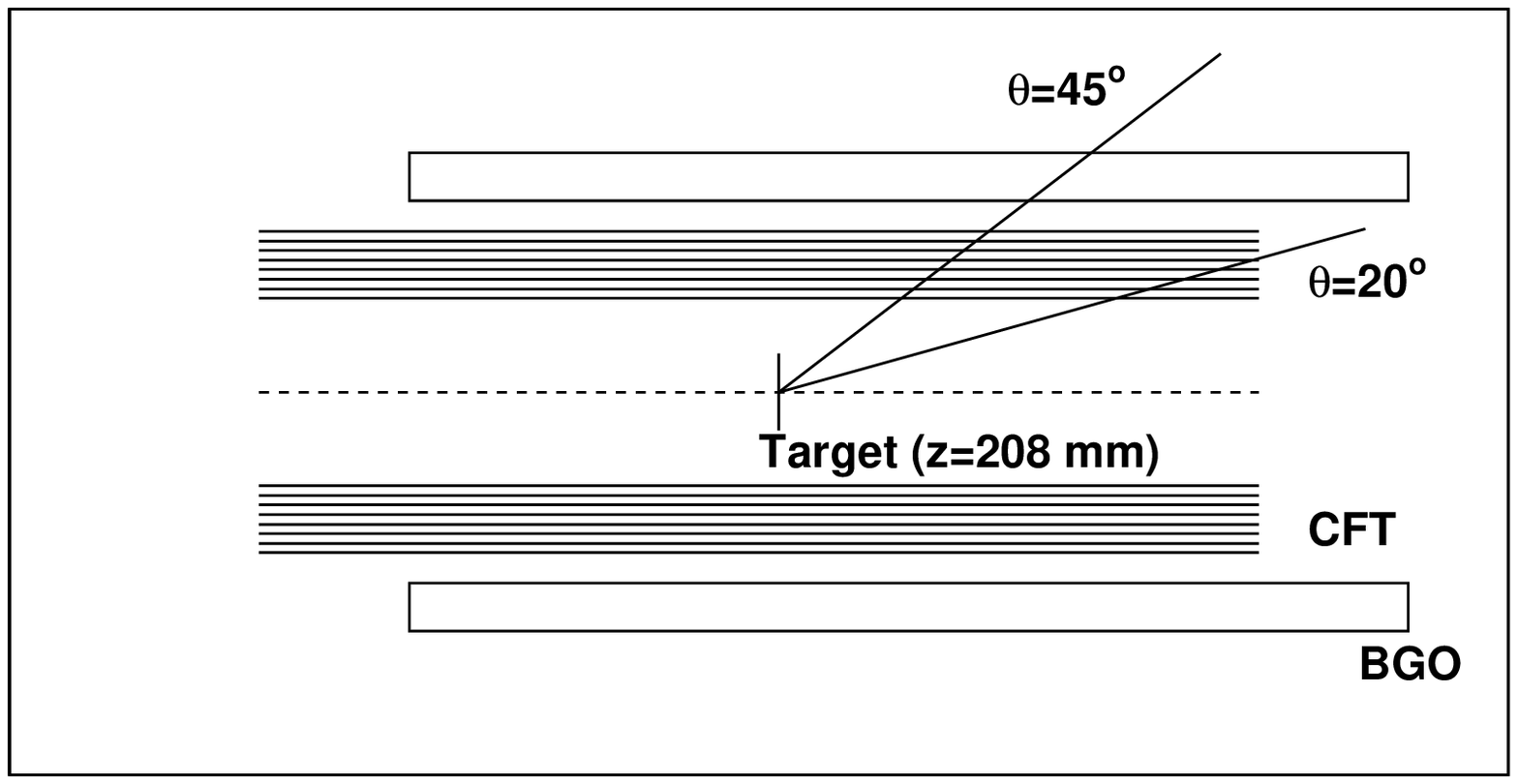}
\includegraphics[width=70mm]{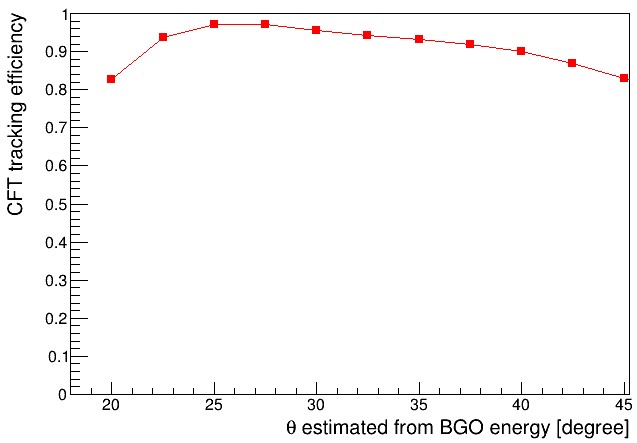}
\caption[CFT tracking efficiency]{
\color{black}
(Top) 
Schematic drawing of the detector geometry with tracks of scattering angles of 20$^{\circ}$ and 45$^{\circ}$.
The forward angular coverage of 20$^{\circ}$ is determined by the CFT's acceptance. 
\color{black}
(Bottom) $\theta$ dependence of the CFT tracking efficiency in which $\theta$ is estimated through the energy of protons measured using the BGO calorimeter. 
}
\label{effCFTtracking}
\end{figure}

\begin{figure}
\centering
\includegraphics[width=60mm]{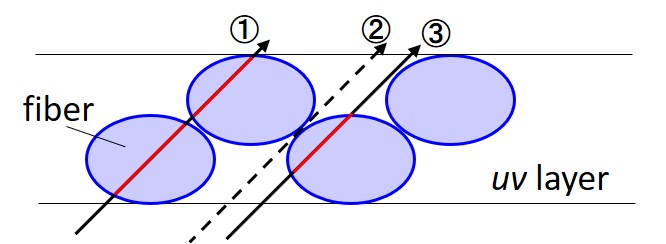}
\caption[Path length in fibers of a CFT spiral layer]{
\color{black}
Cross section of a CFT $uv$ layer and the position dependence of the pass length in the layer.
The cross section becomes an ellipse with respect to a plane including the beam axis owing to the tilt angle in the layer. 
There are ineffective regions for the tracks with scattering angles at approximately 45$^{\circ}$ owing to the zigzag structure of the $uv$ layers.
The "track 2" shown by the dotted line is a typical example of such undetectable tracks.
\color{black}
}
\label{path_structure}
\end{figure}

 The CFT tracking efficiency depends on both the kinetic energy and the scattering angle of the proton.
However, because there is a one-to-one correlation between the kinetic energy and the scattering angle of the $pp$ elastic scattering, we estimate the CFT efficiency as a function of the scattering angle for the $pp$ elastic scattering.
This efficiency is necessary for the derivation of the differential cross section, as described in the next section.
To select the $pp$ scattering events without the kinematic identification, we required two fiber hits with a back-to-back correlation in the $\phi 1$ layer corresponding to the two protons from the $pp$ scattering experiment. 
In this analysis, the scattering angles were estimated from the BGO energy because there is a one-to-one correlation between them in the $pp$ scattering experiment, as shown in Figure \ref{bgo_pp}.
The tracking efficiency for a scattering angle $\theta$ was estimated as follows:
\begin{equation} 
\epsilon_{CFT}(\theta) = \frac{N_{CFT}(E_{BGO})}{N_{BGO}(E_{BGO})}, \label{equ_CFT_eff}
\end{equation} 
where $N_{BGO}(E_{BGO})$ and  $N_{CFT}(E_{BGO})$ represent the number of protons detected using the BGO calorimeter and the number of tracked events using the CFT, respectively, for each BGO energy region ($E_{BGO}$) corresponding to the scattering angle $\theta$ based on the $pp$ scattering kinematics.
\color{black} The bottom figure of \color{black} Figure \ref{effCFTtracking} shows the angular dependence of the tracking efficiency of the CFT. 
\color{black}
To estimate $N_{BGO}(E_{BGO})$ in Equation (\ref{equ_CFT_eff}), protons have to arrive at the BGO calorimeter with a reasonable kinetic energy.
Therefore, the estimation of the CFT tracking efficiency is restricted to the angular region less than 45$^{\circ}$.
\color{black}
\color{black}
As shown in the top figure of Figure \ref{effCFTtracking}, the forward angular acceptance is determined by the CFT's acceptance.
The efficiency suddenly drops at the edge region around 20$^{\circ}$.
\color{black}
On the other hand, the efficiency gradually decreases as the scattering angle moves toward 45$^{\circ}$.
This is caused by the geometrical \color{black} effect \color{black} in the $uv$ layers. 
As shown in Figure \ref{path_structure}, 
\color{black}
there are ineffective regions for the tracks with scattering angles at approximately 45$^{\circ}$ owing to the zigzag structure of the $uv$ layers.
\color{black}
\color{black}
It is difficult to detect protons passing near the boundary region of the neighboring fibers, such as "track 2" shown by the dotted line in Figure \ref{path_structure}.
\color{black}
This geometrical effect reaches the maximum at 45$^{\circ}$. 
Therefore, the efficiency gradually drops as the scattering angle approaches 45$^{\circ}$.



 \subsection{Relative differential cross sections of the $pp$ elastic scattering experiments}
\begin{figure}
\centering
\includegraphics[width=80mm]{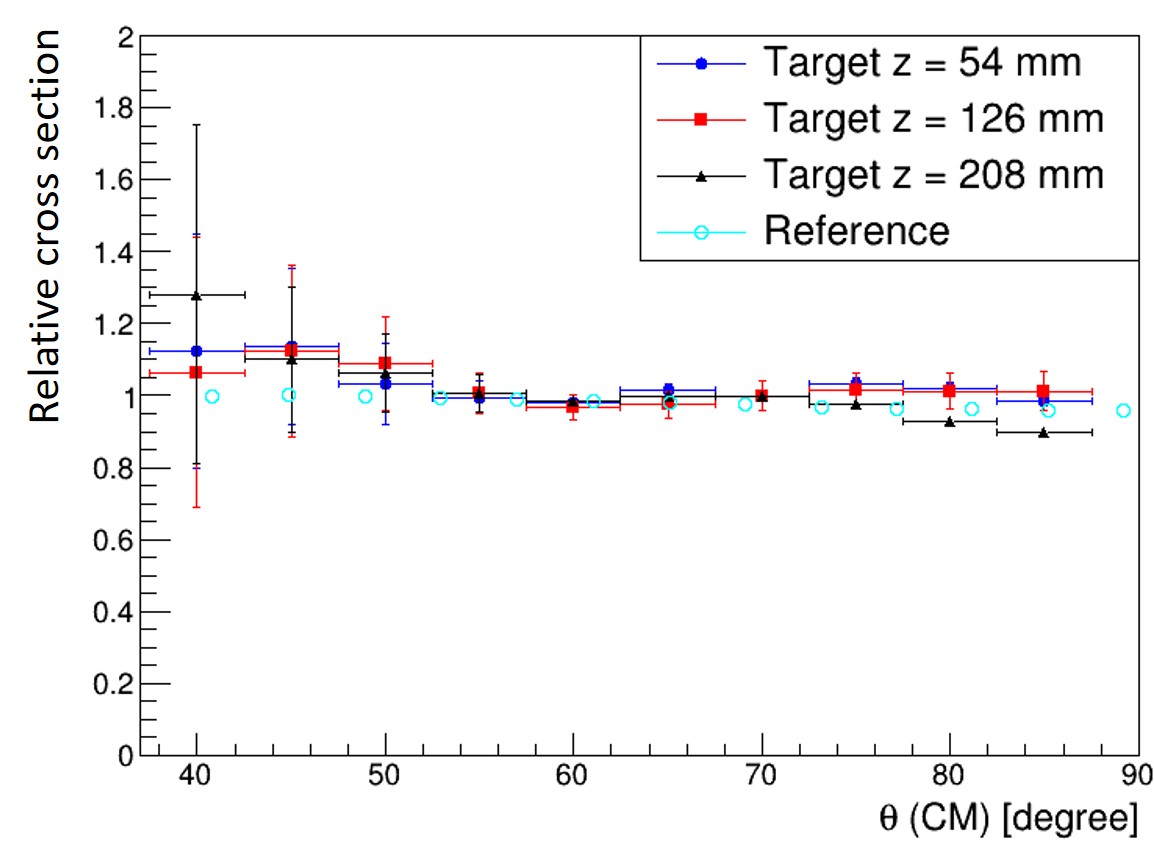}
\caption[Relative cross section of the $pp$ and $p$C scatterings]{
Relative differential cross sections, which are normalized to one at $\theta_{CM}$=70$^{\rm o}$, for the $pp$ elastic scatterings obtained in this experiment for three different target positions.
The results obtained from a partial-wave analysis based on rich experimental data of nucleon-nucleon scattering experiments \cite{Arndt:2007} is also shown as a reference.
}
\label{cross}
\end{figure}

 To evaluate the performance for deriving the differential cross sections using CATCH, we finally derived the differential cross section of the $pp$ scattering experiment. The differential cross section was derived as follows:
 \begin{equation}
\frac{d\sigma}{d\Omega} = \frac{N_{scat}(\theta_{CM})}{N_{beam}\cdot N_{target} \cdot \epsilon_{CFT} \cdot \epsilon_{acc}\cdot d\Omega},
 \end{equation}
 where $N_{beam}$, $N_{target}$, and $d\Omega$ represent the number of proton beams, the number of the target protons per unit square, and the solid angles, respectively. $\epsilon_{CFT}$ and $\epsilon_{acc}$ represent the tracking efficiency of the CFT and the detector acceptance representing the probability of arriving at the BGO calorimeter, respectively. The acceptance $\epsilon_{acc}$ was obtained based on the Monte Carlo simulation.
$N_{scat} (\theta_{CM})$ represents the number of the protons scattered to the scattering angle $\theta_{CM}$ in the c.m. frame. This $N_{scat} (\theta_{CM})$ is obtained from the $\Delta E_{kin}$ spectrum for each scattering angle. In this experiment, the $N_{beam}$ could not be measured accurately because the intensity of the beam was low.
Therefore, it is difficult to derive the absolute value of the differential cross section. Here, we discuss the relative differential cross section by normalizing the differential cross section at $\theta_{CM}$=70$^{\rm o}$ to one.
Figure \ref{cross} shows the obtained relative differential cross section for the angular region of 40 $\le$ $\theta_{CM}$ (degree) $\le$ 90. The differential cross section was derived for the three different target positions to determine the $z$ dependence of the detector performance, and the obtained values were almost consistent with each other. The obtained relative differential cross section was almost flat in this angular region. 
A result obtained from a partial-wave analysis \cite{Arndt:2007} based on the rich experimental data of nucleon-nucleon scattering experiments is also referred to as the cyan circle, as shown in Figure \ref{cross}.
We concluded that the obtained relative differential cross section was reasonably consistent with the reliable calculation. However, there are significant errors, especially regarding both edges of the angular acceptance.
This error in such angular regions results from the rapid drop in the total efficiency ($\epsilon_{CFT}\cdot \epsilon_{acc}$). 
\color{black}
Even with discrepancies in the cross-sections for the different target positions, the systematic errors are below 10\%.
\color{black}
The systematic error mainly results from the uncertainty in the efficiency estimation of the CFT, especially at the edge of the angular acceptance.
This estimation can be improved using a higher-energy proton beam because two proton tracks can be reconstructed using CATCH, whereas two-track reconstruction is difficult in this experiment owing to the low energy.
\color{black}
Detecting a proton from the $pp$ scattering, one can reconstruct the other in the $pp \to pX$ reaction's missing momentum.
The tracking efficiency is defined as the number of detected tracks divided by the number of predicted ones.
\color{black}
In our future studies, we shall improve efficiency estimation using this method.
The current systematic error also satisfies the requirements because the statistical uncertainty in $\Sigma p$ scattering experiments is expected to be at 10\%.
Therefore we concluded that CATCH \color{black} meets \color{black}  
the required performance for deriving the differential cross section.

 \section{Summary}

In summary, we completed the development of a recoil proton detector called CATCH for the $\Sigma p$ scattering experiment at J-PARC.
CATCH is a cylindrical detector system covering an inner target to measure the trajectory and energy of an emitted proton from the target for the kinematic identification of the $\Sigma p$ scattering event.
CATCH comprises a cylindrical fiber tracker (CFT), a BGO calorimeter, and a plastic scintillator hodoscope (PiID), which are coaxially arranged from the inner to the outer sides. The CFT is the tracking detector comprising eight cylindrical layers of scintillation fibers arranged in parallel and spiral configurations based on the direction of the beam to obtain a three-dimensional trajectory.  
 Each CFT fiber is read out one after the other using an MPPC, and approximately 5,000 MPPCs are controlled using the VME-EASIROC boards, which are readout boards using the EASIROC chips designed for multi-PPD readout. As for the BGO calorimeter, $24$ BGO crystals are placed around the CFT. To operate the BGO calorimeter under the high singles rate condition, the readout PMT is operated at a lower voltage, and the waveform of the shaping amplifier signal is read out using a flash ADC. Because the fiber positions were slightly deviated from the desired positions, the fiber positions were corrected in the offline analysis using cosmic ray data. As a result of the fiber position correction, the angular resolution of the CFT was 1.30$^{\circ}$ in $\sigma$ for the cosmic ray.

To evaluate the performance of CATCH for the protons, the $pp$ and $p$C scattering experiments were performed at the cyclotron facility in Tohoku University (CYRIC) in January 2017. An 80 MeV proton beam was irradiated on a thin polyethylene (CH$_2$) target installed inside CATCH. The scattered protons were detected using CATCH. The angular resolution and the time resolution of the CFT for protons were 1.27$^{\circ}$ and 1.8 ns in $\sigma$, respectively. The energy calibrations for the BGO calorimeter and the CFT were performed by comparing them to the simulated energy deposits for each scattering angle. Although the CFT and the BGO calorimeter were calibrated separately, the reconstructed correlation between the scattering angle and the kinetic energy of the scattered protons was well consistent with the kinematic line. To identify the $pp$ scattering events, the index $\Delta E_{kin}$ corresponding to the kinematic consistency between the measured energy and the angle was introduced. 
 The $\Delta E_{kin}$ resolution was 2.8 MeV in $\sigma$ for the $pp$ scattering of the entire $\theta$ region, and the $pp$ scattering event was identified by selecting the peak region in the $\Delta E_{kin}$ distribution. The obtained relative differential cross sections for the $pp$ elastic scatterings were consistent with the reliable partial wave analysis for three target positions. The systematic error was maintained at under 10\%, which is the required performance level for CATCH. 
From these performance evaluations, we confirmed that CATCH demonstrates sufficient performance for the $\Sigma p$ scattering experiment at J-PARC. 

\section*{CRediT authorship contribution statement}
{\bf Y. Akazawa:} Supervision, Software, Formal analysis, Investigation, Writing - Original Draft, Visualization, 
{\bf N. Chiga:} Methodology,
{\bf N. Fujioka:} Methodology, Investigation, 
{\bf S.H. Hayakawa:} Software, Investigation, Writing - Review \& Editing,
{\bf R. Honda:} Methodology, Software, Investigation, Writing - Review \& Editing,
{\bf M. Ikeda:} Methodology, Software, Investigation, 
{\bf K. Matsuda:} Investigation,
{\bf K. Miwa:} Project administration, Conceptualization, Software, Formal analysis, Investigation, Writing - Original Draft, Funding acquisition, 
{\bf Y. Nakada:} Methodology, Software, Investigation, 
{\bf T. Nanamura:} Investigation,
{\bf S. Ozawa:} Methodology, Software, Investigation, 
{\bf T. Shiozaki:} Methodology, Software, Investigation, 
{\bf H. Tamura:} Funding acquisition, 
{\bf H. Umetsu:} Methodology.

\section*{Declaration of Competing Interest}
The authors declare that they have no known competing financial interests or personal relationships that could have appeared to influence the work reported in this paper.

\section*{Acknowledgements}
We thank the staff of CYRIC for their support in providing the beam used for the experiment.
We are grateful to  S. Callier and the OMEGA group for arranging for the EASIROC chip.
We also thank the KEK electronics group for developing the VME-EASIROC board. 
This study was supported by JSPS KAKENHI Grant Numbers \\
 23684011, 15H05442, 15H02079, and 18H03693. 
It was also supported by Grants-in-Aid Numbers 24105003 and 18H05403 for Scientific Research from the Ministry of Education, Culture, Science, and Technology (MEXT), Japan.

\bibliography{CATCH_NIMA}

\end{document}